\documentclass[journal]{IEEEtran}
\usepackage[utf8x]{inputenc}
\usepackage[english]{babel}
\usepackage{verbatim} 
\usepackage{varioref}
\usepackage{subfigure} 
 
\usepackage{psfrag}
\usepackage{epsf}
\usepackage{graphics}
\usepackage{graphicx}
\usepackage{amsbsy}
\usepackage{amssymb}
\usepackage{amscd}
\usepackage{amsmath}
\usepackage{amsthm}
\usepackage{enumerate} 
\usepackage{color}
\usepackage{epstopdf}

\usepackage{threeparttable}
\usepackage{algorithm2e}

\usepackage{cite}
\usepackage{url}

\DeclareMathOperator{\E}{E}

\DeclareMathOperator{\trace}{Tr}

\DeclareMathOperator*{\argmax}{argmax}

\newtheorem{remark}{Remark}

\def\mathC{{\mathbb{C}}}

\def\bH{{\boldsymbol{H}}}

\title{How Much Spectrum is Too Much in Millimeter Wave Wireless Access}

\author{ 
Jinfeng Du, \emph{Member, IEEE}, and Reinaldo A. Valenzuela, \emph{Fellow, IEEE}
\thanks{Authors are with Nokia Bell Labs, Holmdel, NJ 07733, USA. Email: 
\{jinfeng.du, rav\}@bell-labs.com. 
}} 

\begin{document}

\maketitle

\begin{abstract}

Great increase in wireless access rates might be attainable using the large amount of spectrum available in the millimeter wave (mmWave) band. However, higher propagation losses inherent in these frequencies must 
be addressed, especially at ranges beyond 100 meters and in non-line-of-sight (NLOS) settings. 
In contrast to the interference limited legacy cellular systems where using more bandwidth is favorable, , to use wider bandwidth for mmWave channels in noise limited settings may be ineffective  or even counterproductive when accounting for channel estimation penalty. 
In this paper we quantify the maximum beneficial bandwidth for mmWave transmission in some typical deployment scenarios where pilot-based channel estimation penalty is taken into account assuming a minimum mean square error (MMSE) channel estimator at the receiver. 
We find that, under I.I.D. block fading model with coherence time $T_c$ and coherence bandwidth $B_c$, 
for transmitters and receivers equipped with a single antenna, the optimal (rate maximizing) signal-to-noise-ratio (SNR) is a constant that only depends on the product $B_cT_c$, which measures the channel coherence and equals the average  number of orthogonal symbols per each independent channel coefficient. 
That is, for fixed channel coherence $B_cT_c$, the optimal bandwidth scales linearly with the received signal power.  
Under 3GPP Urban Micro NLOS path loss model with  coherence time $T_c=5$ ms and coherence bandwidth $B_c=10$ MHz, using 52 dBm Equivalent  Isotropic Radiated Power (EIRP) at the transmitter and 11 dBi antenna gain at the receiver,    the maximum beneficial bandwidth at 28 (resp. 39) GHz is less than 1 GHz at a distance beyond 210 (resp. 170) meters with maximum throughput about 200  Mbps, and  less than 100 MHz beyond 400 (resp. 310) meters with maximum throughput about 20~Mbps.  
At  EIRP of 85 dBm, corresponding to the FCC limit of 75 dBm per 100 MHz,  1 Gbps rate can be delivered using 1 GHz bandwidth up to 860 (resp. 680)~meters.
\end{abstract}
 

\section{Introduction}\label{sec:introduction}

Spectrum available in the millimeter wave (mmWave) band is expected to be tens of GHz as compared to hundreds of MHz in legacy cellular band. For example, in the United States, the FCC~\cite{FCC2016} recently has opened up 3.85 GHz of licensed spectrum with an Equivalent  Isotropic Radiated Power (EIRP) limit of 75 dBm for every 100 MHz bandwidth at base stations\footnote{A peak EIRP limit of 43 dBm is proposed for  {mobile stations} and 55 dBm for \emph{transportable stations}, e.g., modem.}, and 14 GHz of unlicensed spectrum (57 GHz--71 GHz) with 40 dBm EIRP limit\footnote{For outdoor fixed point-to-point backhaul links with high antenna gains, the EIRP limit is up to 82 dBm.} to facilitate wireless broadband and next generation wireless technologies in spectrum above 14 GHz. The abundant spectrum in mmWave has attracted a lot of efforts both in academia and in industry to measure and model mmWave channels~\cite{mustafa2013mmWave, Amitava2014, NYU2015, 5GCM-VzW, 3GPP-above6G} and to evaluate its potential for future wireless systems. See also~\cite{Andrews2016} for a rich literature survey of advances in mmWave.
 
However, higher propagation losses inherent in mmWave frequencies must be addressed, especially at ranges beyond 100 meters and in non-line-of-sight (NLOS) settings. On one hand, higher propagation loss and higher foliage loss make it challenging to deliver signals. On the other hand, the limited transmit signal power spreading over a large bandwidth may drive the link into the low signal to noise ratio (SNR) regime. Therefore, in contrast to the interference limited legacy cellular systems where using more bandwidth is favorable, in mmWave systems, users located at range of hundred meters and beyond are most likely to be noise limited, and using more bandwidth leads to a diminishing return and can even be   counterproductive when the channel estimation penalty exceeds the gain of increasing the degree-of-freedom (DoF). 

From information theoretical point of view, it is known that using too much bandwidth might be counterproductive in wideband fading channels, unless the signal becomes extremely peaky, in the sense that its fourth moment grows at least as fast as signaling bandwidth~\cite{journals/tit/MedardG02}. That is,   concentrating the power into a vanishing fraction of its symbols and transmitting infrequent bursts, as shown in~\cite{Telatar-Tse2000} using a peaky Frequency-Shift-Keying modulation with a vanishing duty-cycle to facilitate non-coherent detection at the receiver. 
Capacity analysis  with both peak and average power constraints studied in~\cite{Durisi2010} has demonstrated bell-shaped upper and lower bounds that peak at finite bandwidths. For spread spectrum transmission over wideband fading channels using Binary-Phase-Shift-Keying modulation, it was shown~\cite{MedardThesis, MedardTse} that there exists a maximum beneficial bandwidth spreading, beyond which over-spreading will decrease the rate.  
For single-input single-output (SISO) wideband fading channels, mutual information based analysis in~\cite{Lozano-Porrat} has provided closed-form  bounds to coarsely identify the range within which the ``critical bandwidth'' locates. This analysis has been generalized in~\cite{GDME2015} to multiple-input multiple-output (MIMO) systems accommodating duty cycles and a set of new tools are developed to bound the optimal bandwidth\footnote{More precisely, in~\cite{GDME2015} it refers to the ``bandwidth occupancy'' that measures the average bandwidth usage over time.}.

In practical mmWave systems, channel state information (CSI) is crucial to  materialize the large beamforming gain out of the many-element antenna array to sustain a feasible link budget~\cite{CVV2014,CVV2015,CVVW2013}.
The associated channel estimation cost grows with the number of antennas\footnote{If each receive antenna is equipped with an independent RF-chain, the channel estimation cost grows only with the number of transmit antennas.} and the signaling bandwidth. Rate penalty caused by channel estimation overhead and channel estimation error has been analyzed in~\cite{Hassibi-Hochwald} for the pilot-based minimum mean square error (MMSE) channel estimator. The cost and quality of channel estimation has been investigated in~\cite{CVV2015} for a feedback based  beam-switching system, where the fraction of pilot is optimized to maximize the rate. 

In this work, we focus on mmWave transmission systems that exploit pilots for channel estimation and we are interested in determining the optimal signal bandwidth and fraction of pilots for channel estimation to maximize user throughput at a range of power levels and link distances in some typical deployment scenarios.   
Our investigation has revealed a surprising result that, under I.I.D. block fading model with coherence time $T_c$ and coherence bandwidth $B_c$, for transmitters and receivers equipped with a single antenna, the optimal (rate maximizing)  SNR and the optimal pilot ratio are two constants that only depend on the product $B_cT_c$, which measures the channel coherence and equals the average  number of orthogonal symbols per each independent channel coefficient. That is, for fixed channel coherence $B_cT_c$, the optimal bandwidth scales linearly with the received signal power, and the percentage of pilots is fixed whereas the absolute number of pilots grows linearly. Our analysis is then extended to beamforming based MIMO systems where a single data stream is transmitted (per each polarization\footnote{Throughout this paper we only use a single polarization for analysis. It is common in practice to have separate RF power supply chain for each polarization, in which case we can duplicate the analysis to fit the setup.}).
Furthermore, we find that, under 3GPP Urban Micro NLOS path loss model~\cite{3GPP-below6G} with  coherence time $T_c=5$ ms and coherence bandwidth $B_c=10$ MHz, using 52 dBm EIRP at the transmitter and 11 dBi antenna gain at the receiver, the maximum beneficial bandwidth at 28/39/60 GHz band is less than 1~GHz for users beyond 210/170/120~meters with corresponding rate at around 200~Mbps. The maximum beneficial bandwidth is less than 100~MHz when the transmission distance is longer than 400/310/230~meters. 

The rest of this paper is organized as follows. We present the system model in Sec.~\ref{sec:model} and derive the bandwidth-pilot optimization problem and its solutions for  SISO systems in Sec.~\ref{sec:Opt-siso}. The results are extended to beamforming based MIMO systems in Sec.~\ref{sec:Opt-mimo} and the maximum beneficial bandwidth is evaluated in Sec.~\ref{sec:num} for the 28, 39, and 60 GHz bands with different power level and link parameters. Two potential use cases are presented in Sec.~\ref{sec:use}, and conclusions and future work are in Sec.~\ref{sec:conclusion}.
 
\section{System Model}\label{sec:model}
 
First we focus on the case where both the transmitter and the receiver are equipped with a single antenna, formulating and solving the joint bandwidth and pilot optimization problem for the SISO case in Sec.~\ref{sec:Opt-siso}, and then extend the analysis to beamforming based MIMO systems in Sec.~\ref{sec:Opt-mimo}.

\subsection{Channel Model and Channel Coherence}\label{sec:channel} 
  
We adopt the discrete-time channel model where the channel input-output relation can be written as 
\begin{align}
y=\sqrt{P_r} hx + z, \label{eqn:siso-channel}
\end{align}	
where $x\in\mathC$ is the power-normalized complex valued input signal with unit average power constraint $\E[|x|^2]=1$, $h\in\mathC$ is a complex valued random variable with $\E[|h|^2]=1$ that represents the small scale channel fading,  $y\in\mathC$ is the complex-valued channel output signal,  and $z\in\mathC$ is the circularly symmetric complex Gaussian additive noise with power spectral density $N_0$ (Watt/Hz).
Note that the transmit signal power, transmit and receive antenna element  gains, path loss, {vegetation absorption,} and other attenuation factors are incorporated into the received signal power $P_r$ (Watt). With signaling bandwidth $W$ (Hz), the SNR can be written as
\begin{align}
\rho(W)=\frac{P_r}{N_0W}. \label{eqn:siso-SNR}
\end{align}
 
We further assume that the channel coefficient $h$ stays unchanged for a time period equals a coherence time $T_c$ (seconds) and over a bandwidth equals a coherence bandwidth $B_c$ (Hz), after which $h$ will change independently at random. This is often referred as I.I.D. block fading channel model where the product $L_c{\triangleq} B_cT_c$ is defined as the channel coherence length, or channel coherence for short. {For signal symbols of bandwidth $W$ and symbol duration $T\geq \frac{1}{W}$, the number of  symbols that can be transmitted over a coherence block of time $T_c$ and  bandwidth $B_c$ can be written as
\[\frac{B_c}{W}\frac{T_c}{T} = \frac{B_cT_c}{WT} \leq B_cT_c, \]
 with the equality holds when $T=1/W$. Therefore} the channel coherence length $L_c$ quantifies the average number of independent\footnote{{This number can be reduced by adding guarding time/frequency between a  group of packed symbols to reduce interference. The number can also be increased but at the cost of losing independence among symbols.}} signal symbols that can be packed over the time-frequency resource block of $T_c\times B_c$.

\subsection{Pilot Based Channel Estimation, Data Transmission, and Detection}\label{sec:estimation}

We adopt the pilot-based MMSE channel estimator and its associated channel estimation error analysis~\cite{Hassibi-Hochwald} to facilitate our analysis. Among the $L_c$ symbols transmitted over each coherent time-frequency block, a fraction of them, $\alpha L_c$ with $\alpha{\in(0, 1)}$,  are used as pilots for channel estimation and the remaining $(1-\alpha)L_c$ symbols are used to transmit data. The MMSE channel estimation $\hat{h}$ and its associated channel estimation error $\tilde{h}$ can be characterized~\cite{Hassibi-Hochwald} as\footnote{
We start with the setup where both pilot symbols and data symbols use the same transmit power. Different power allocation for pilots and data symbols will be discussed in Sec.~\ref{sec:pilot_power}.}
\begin{align}
& h=\hat{h} + \tilde{h}, \ \ \E[\hat{h}\tilde{h}]=0, \label{eqn:h-estimate1}\\
& \E[|\hat{h}|^2]=\frac{\alpha L_c \rho(W)}{1+\alpha L_c \rho(W)}, \ \E[|\tilde{h}|^2]=\frac{1}{1+\alpha L_c \rho(W)}. \label{eqn:h-estimate}
\end{align}

The receiver, after performing the MMSE channel estimation based on the known pilots, treats the estimated channel  $\hat{h}$  as the true channel and discards the signal associated with the estimation error $\tilde{h}$ as noise. Therefore the channel capacity lower bound, highlighted with the signaling bandwidth $W$ and pilot ratio $\alpha$, can be written~\cite{Hassibi-Hochwald} as ({bits per second, bps})
\begin{align}
R(W, \alpha) = (1-\alpha)W\E[\log_2(1+\rho_{\mbox{eff}}(W,\alpha)|h|^2)],  \label{eqn:Rwa}
\end{align}
where the effective SNR $\rho_{\mbox{eff}}(W,\alpha)$ is given as %  
\begin{align}
\rho_{\mbox{eff}}(W,\alpha) & \triangleq \frac{\rho(W)\E[|\hat{h}|^2]}{1 + \rho(W)\E[|\tilde{h}|^2]} 
= \frac{\alpha L_c \rho(W)^2}{1+\alpha L_c \rho(W) +\rho(W)} \nonumber\\
& = \frac{\alpha L_c P_r^2/N_0^2}{ W^2 + (1+\alpha L_c)WP_r/N_0}, \label{eqn:SNReff}
\end{align}
{where the first step is obtained by treating the estimated channel $\hat{h}$  as the true channel and  the signal associated with MMSE channel estimation error as noise, the second step is obtained by substituting \eqref{eqn:h-estimate}, and the last step comes from \eqref{eqn:siso-SNR}.} 

\section{Joint Optimization of Bandwidth and Pilots for SISO}\label{sec:Opt-siso}

The rate in \eqref{eqn:Rwa} is the ergodic capacity of a fading channel with 
effective SNR $\rho_{\mbox{eff}}(W,\alpha)$  given in \eqref{eqn:SNReff}, and the expectation is taken over all channel realizations of $h$ by coding across time-frequency coherence blocks to average out the less favorable channel fading realizations. For any given channel fading distribution, we seek to find the  signaling bandwidth $W^*$ and pilot ratio $\alpha^*$ to maximize the achievable rate $R(W,\alpha)$, that is, 
\begin{align}
(W^*, \alpha^*) = \argmax_{W, \ \alpha} R(W,\alpha). \label{eqn:Opt_orig}
\end{align}
Given our modeling of I.I.D. block fading channel with coherence bandwidth $B_c$ and coherence time $T_c$ (i.e., channel coherence  $L_c=B_cT_c$), both $W$ and $\alpha$ in the optimization \eqref{eqn:Opt_orig} are discrete: $W=mB_c$ and $\alpha=n/L_c$ with $m,n\in\mathbb{N}$ and $n\leq L_c{-}1$. 

Instead of solving the discrete optimization problem via two-dimension exhaustive search, which has to be done for each realization of the triplet $(P_r/N_0, B_c, T_c)$, we first relax $W>0$ and $\alpha\in(0,1)$ to be continuous valued and solve the joint optimization analytically by taking partial derivation of $R$ with respect to $W$ and $\alpha$, respectively. Once we get the continuous valued solution $(W^*, \alpha^*)$, we   map it to the closest discrete point in the two dimensional plane specified by $(W,\alpha)$.  
In the remaining part of this section, we will focus on the continuous valued relaxed optimization problem itself and present the numerical solution in Sec.~\ref{sec:relax_num} and its closed-form approximation in Sec.~\ref{sec:approx}.  
 
\subsection{Solution for the Continuous-Value Relaxed Optimization}\label{sec:relax_num}
We first take partial derivation of $R$ with respect to $W$ and $\alpha$, and set them to zero, which leads to (see Appendix~\ref{app:opt_relax} for the detailed derivation)
\begin{align}
& \E\left[\log(1+ \frac{\alpha L_c\rho^2|h|^2}{ 1 + (1+\alpha L_c)\rho})\right] \label{eqn:partial-D1}\\
&= \frac{2+(1+\alpha L_c)\rho}{1 + (1+\alpha L_c)\rho} - \E\left[\frac{2+(1+\alpha L_c)\rho}{1 + (1+\alpha L_c)\rho + \alpha L_c \rho^2 |h|^2} \right],  \nonumber\\
&\rho(\alpha^2L_c + 2\alpha -1)  =1- 3\alpha,\label{eqn:partial-D2} 
\end{align}
where $\rho=P_r/(N_0W)$ as defined in \eqref{eqn:siso-SNR} is the SNR when using bandwidth $W$. 
 From \eqref{eqn:partial-D1} and \eqref{eqn:partial-D2}, we can determine $(\rho^*, \alpha^*)$ numerically using the algorithms developed in Appendix~\ref{app:opt_num}.
 
We observe that:
\begin{enumerate}
	\item The maximum throughput depends on $W^*$ only through the SNR $\rho^*\triangleq P_r/(N_0W^*)$, i.e., $W^*$ is proportional to $P_r/N_0$;
	\item $\alpha^*$ and $\rho^*$ retain the same dependence prescribed by {the necessary condition} \eqref{eqn:partial-D2} regardless the distribution of $h$ (as long as the continuous assumption holds);
	\item $\alpha^*$ and  $\rho^*$ depend only on $L_c$ and the probability distribution function (PDF) of the channel fading $p_h(|h|)$; that is, for given $L_c$ and channel fading PDF $p_h(|h|)$, the joint bandwidth and pilot optimization will allocate the amount of bandwidth proportional to $P_r/N_0$ to maintain a constant SNR $\rho^*$.  
\end{enumerate}

The implication of the above observation is striking: if there is sufficient amount of spectrum, {as is the case for mmWave band,} fixed spectral efficiency transmission (coding/modulation/pilots determined by $\rho^*$ and $\alpha^*$) is sufficient to maximize the pilot-based transmission throughput. For given $L_c$ and  channel fading PDF $p_h(|h|)$,  only  the signaling bandwidth $W$ needs to be adapted and it grows linearly with the received signal power.

\subsection{Closed-Form Approximation for the Throughput Maximizing $(W^*,\alpha^*)$}\label{sec:approx}

{Since the channel coherence length $L_c$ measures the number of independent symbols that can be placed over each time-frequency resource block of $B_c\times T_c$, it is a quantitative measure that is inversely proportional to the ``speed'' of channel variation. The larger $L_c$ is, the ``slower'' the channel changes in the two dimensional time-frequency plane.}
As channel coherence $L_c$ goes to infinity, the channel estimation penalty becomes negligible and therefore we have  $\alpha^*\to 0$ and $W^*\to\infty$ {(hence SNR goes to zero)}. In this section we will quantify the convergence speed by deriving  closed-form approximations of $(W^*,\alpha^*)$ to shed  some insights into the design and operation of the pilot-based communication over wideband fading channels.

\begin{table*}
\centering
		\caption{Comparison of the rates of different transmission schemes}
	\label{tab:rate_comp}
		\begin{tabular}{|c|c|c|c|c|c|}
		\hline
		scheme & pilot & bandwidth & SNR & rate $R$  &rate penalty  $\frac{R_{CSIR}-R}{R_{CSIR}}$ \\ \hline
		peaky FSK\cite{Telatar-Tse2000} & No & $\infty$ & 0 & \eqref{eqn:Rate_FSK}  & $\frac{1}{L_c}$ \\ \hline
		non-peaky Mutual Information\cite{Lozano-Porrat} & -- &  $O(\frac{P_r}{N_0}\sqrt{\frac{L_c}{\log(L_c)}})$ 		&  $O(\sqrt{\frac{\log(L_c)}{L_c}})$ & \eqref{eqn:Rate_MI} & $O(\sqrt{\frac{\log(L_c)}{L_c}})$ \\	\hline
    non-peaky pilot-based $(\rho^*,\alpha^*)$ & Yes &  $O(\frac{P_r}{N_0}\sqrt[3]{L_c})$ & $O(\frac{1}{\sqrt[3]{L_c}})$ & \eqref{eqn:rate_opt} & $O(\frac{1}{\sqrt[3]{L_c}})$\\
		\hline
		\end{tabular}
\end{table*}

The closed-form approximation, derived in Appendix~\ref{app:opt_approximation}, can be summarized as follows
\begin{align}
\rho^* & =\frac{P_r}{N_0W^*} = (\frac{4}{L_c})^{1/3} + O( {L_c^{-2/3}}),   
\label{eqn:rho_opt}\\
\alpha^*  & = ({2L_c})^{-1/3} + O( {L_c^{-2/3}}), 
 \label{eqn:alpha_opt}\\  
R(\rho^*,\alpha^*) &= \left(1-(\frac{4}{L_c})^{1/3} +O({L_c^{-2/3}})\right)\frac{P_r}{N_0}\log_2(e). \label{eqn:rate_opt}
\end{align}
 That is,  the throughput maximizing SNR, the pilot ratio, and the rate penalty, all grow inversely proportional to the cubic root\footnote{{The underlining physical explanation for cubic root rather than any other root is yet to be determined.}} of the channel coherence $L_c$.

It is worthwhile to compare different transmission schemes and their corresponding rate penalty with respect to the full-CSIR infinite-bandwidth capacity 
\begin{equation}
R_{CSIR}=\frac{P_r}{N_0}\log_2(e).
\end{equation}
 In \cite{Telatar-Tse2000} a non-coherent transmission scheme has been proposed by combing  infinite-bandwidth FSK modulation with a duty cycle to concentrate the signal power to a small fraction of active FSK symbols. The achievable rate is\footnote{The result in \cite{Telatar-Tse2000} has a rate penalty $2/L_c$ owing to the two guard time intervals added before and after the data symbols. The rate penalty has been improved to $1/L_c$ in \cite{Medard2005} by using only one guard interval.}
\begin{align}
R_{peaky-FSK} = (1-\frac{1}{L_c})\frac{P_r}{N_0}\log_2(e), \label{eqn:Rate_FSK}
\end{align}
 which can only be approached when the duty cycle goes to zero, i.e., using infinite high signal power for an infinitesimal fraction of time to maintain the average power constraint.  
Another benchmark is the mutual information based rate \cite{Lozano-Porrat},
\begin{align}
R_{non-peaky MI} = \left(1-\sqrt{\frac{\kappa\log(\pi)\log(L_c)}{L_c}}\right)\frac{P_r}{N_0}\log_2(e), \label{eqn:Rate_MI}
\end{align}
 where $\kappa$ is the kurtosis of the channel $h$, and the rate is achievable by using non-peaky signaling with a sufficiently large but finite bandwidth. Here the achievability refers to the fact that there exists a coding scheme that achieves the rate calculated based on mutual information, but the explicit transmission/detection scheme (e.g., pilot vs. no pilot) is yet to be determined. A detailed comparison of the three schemes is shown in Table~\ref{tab:rate_comp}. 
{When there is no restriction on how/whether channel estimation is performed, such as the mutual information based study in~\cite{Lozano-Porrat, GDME2015}, the optimal SNR and the associated rate penalty grow inversely proportional to $\sqrt{L_c/\log(L_c)}$. When pilot-based channel estimation is used, as studied in this work, the convergence speed is associated with the cubic root of $L_c$ (with approximation). The speed of the convergence depends on the actual transmission schemes in use and the gap of rates indicates the potential gain of using more advanced transmission schemes such as iterative estimation-detection and joint estimation-detection.}

Note that for fixed signal bandwidth $W$, the spectral efficiency obtained by only optimizing the pilots is shown in~\cite{Hassibi-Hochwald} to be proportional to $\rho^2$ when $\rho{\ll} 1$, which is significantly smaller than the rate \eqref{eqn:rate_opt}. The reason, as explained in \cite{Hassibi-Hochwald}, is that the channel estimation penalty is too large when $\rho{\ll} 1$. By also optimizing the signal bandwidth, we can avoid the too-low-SNR regime and achieve a rate that is close to  capacity. On the other hand, when bandwidth is very limited and therefore the SNR $\rho\gg 1$, the channel estimation penalty is negligible and its spectral efficiency is proportional to $\log(1{+}\rho)$. By adding more bandwidth we can substantially increase the throughput with a lower spectral efficiency. 

\begin{figure} 
	\centering
		\includegraphics[width=0.35\textwidth]{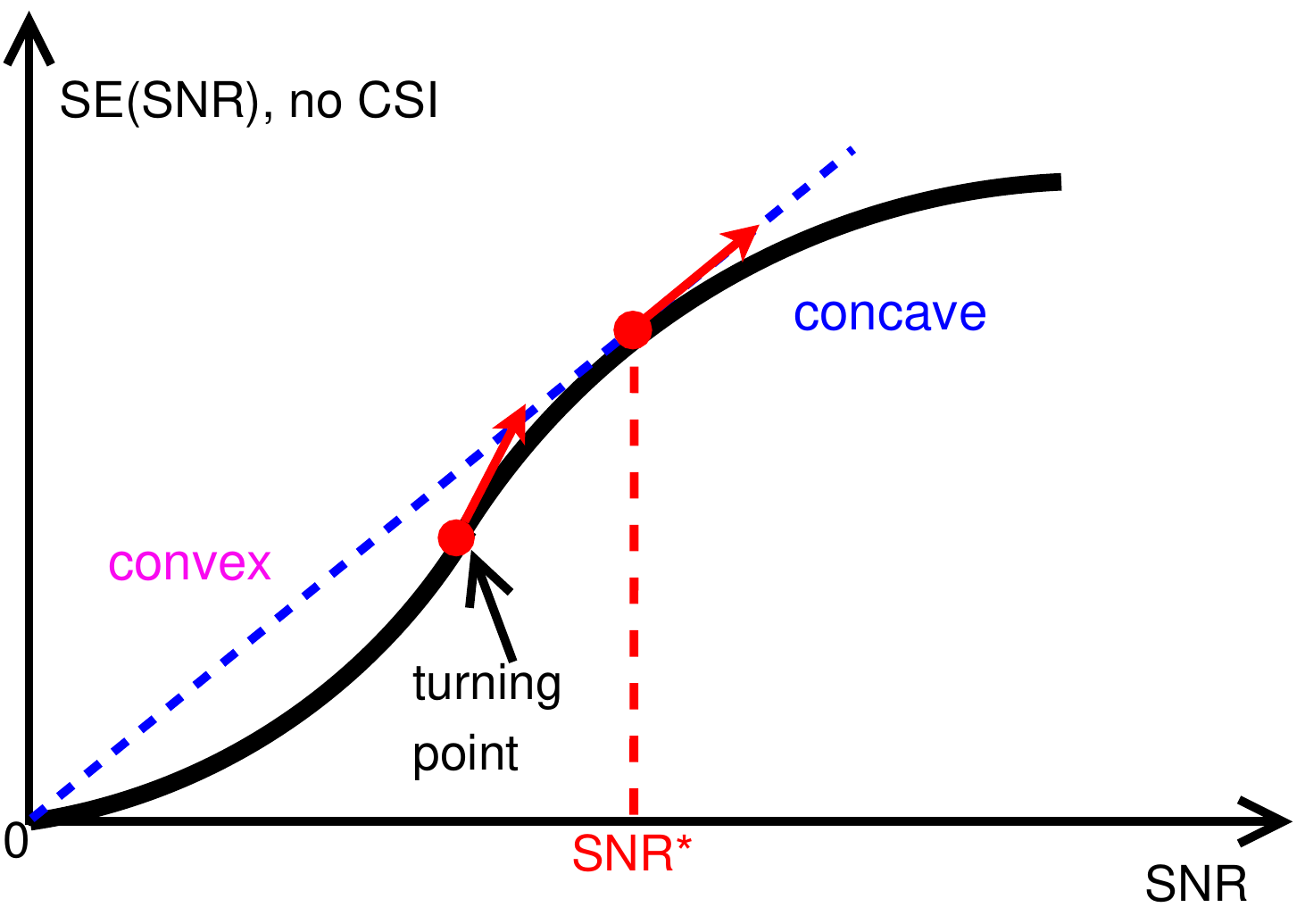}
	\caption{Illustration of the spectral efficiency [bits/s/Hz]  of a wideband fading channel without CSI as a function of the SNR. In the low SNR regime the spectral efficiency grows as $\mbox{SNR}^2$ (hence convex) and in the high SNR regime it grows as $\log(1+\mbox{SNR})$ (hence concave).  
 The maximum rate appears at the SNR where the growth (resp. drop) of spectral efficiency ``matches'' the decrease (resp. increase) of bandwidth ($\propto 1/\mbox{SNR}$). 
The ``convex-concave'' shaped achievable rates and the existence of the optimal SNR are first observed in~\cite[Fig.~II.18]{MedardThesis} (see also \cite[Fig.~1]{MedardTse}) in the analysis of spread spectrum transmission using BPSK modulation over wideband fading channels.
}
	\label{fig:Opt_SNR}
\end{figure}

\begin{figure*} 
	\centering
		\includegraphics[width=\textwidth]{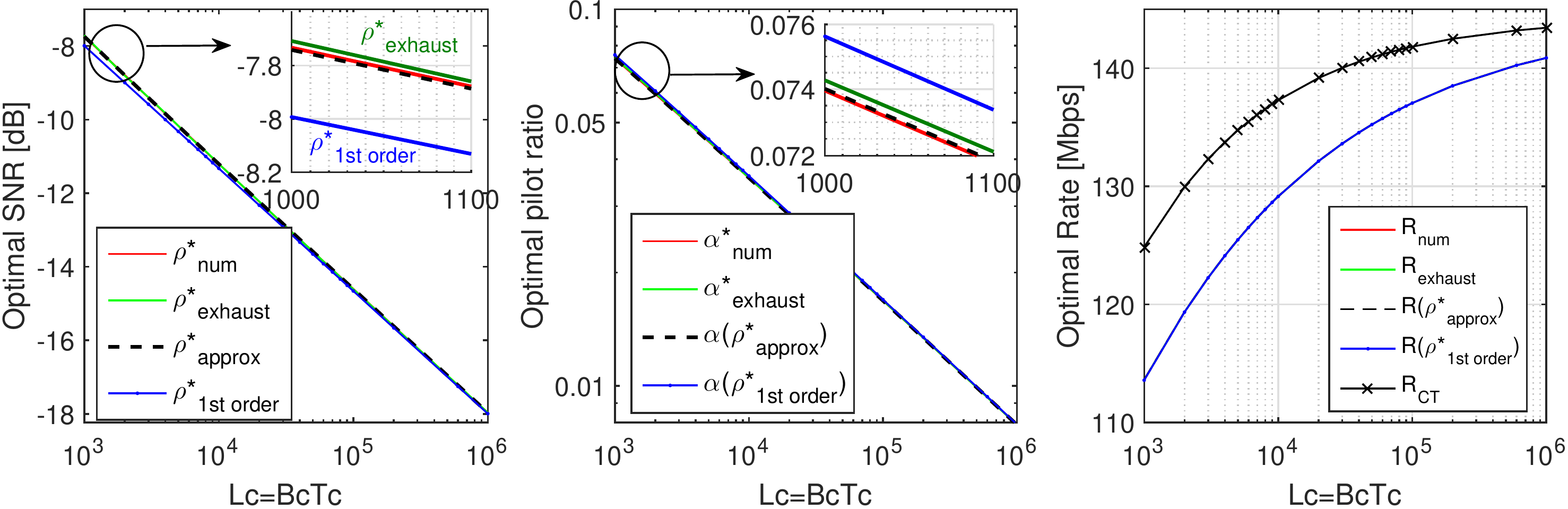}
	\caption{Comparison of the throughput maximizing SNR $\rho^*$, pilot ratio $\alpha^*$, and the corresponding achievable rates obtained by exhaustive search for \eqref{eqn:Opt_orig}, numerical optimization using \eqref{eqn:partial-D2}, the closed-form approximation \eqref{eqn:rho_heur}, and the 1st-order approximation \eqref{eqn:rho_opt}. The non-peaky mutual information \eqref{eqn:Rate_MI} from \cite{Lozano-Porrat} is also plotted for reference. Coherence bandwidth $B_c=10$ MHz and coherence time $T_c\in[0.1, 100]$ ms are used for the simulation,  with $P_r/N_0=20$ dB.MHz (i.e., if $W=1$ MHz is the signal bandwidth, the corresponding SNR is $P_r/(N_0 W)=20$ dB).}
	\label{fig:Opt_rho_alpha_Lc_PrN0_20dB_Bc10MHz}
\end{figure*}

By lifting the constraint on signaling bandwidth, the throughput of pilot-based transmission over wideband fading channels is maximized at a unique SNR, which is not determined by the available (finite) power. Its spectral efficiency (SE) [bits/s/Hz]  as a function of the SNR first grows as  $\mbox{SNR}^2$ (hence convex) in the low SNR regime and in the high SNR regime it grows as  $\log(1{+}\mbox{SNR})$ (hence concave).  
The first order derivative of spectral efficiency  quantifies the ``exchange rate'' of SNR to spectral efficiency. When increase the SNR from zero to infinity, the first order derivative first increases and then decreases, as illustrated in Fig.~\ref{fig:Opt_SNR}, with the maximum exchange ratio peaked at the turning point where its second order derivative equals zero. The maximum rate appears at the SNR where the growth of spectral efficiency ``matches'' the decrease  of bandwidth (${\propto} 1/\mbox{SNR}$), beyond which the bandwidth becomes the limiting factor.

\begin{remark}
To quantify the exchange rate of bandwidth to capacity in the infinite bandwidth regime,  in~\cite{journals/tit/Verdu02} the concept of \emph{wideband slope} was introduced by defining it as inversely proportional to the second order derivative of spectral efficiency at zero SNR, i.e., $\ddot{C}(\rho{=}0)$.  
Here we target at maximizing the data rate for pilot-based transmission schemes, 
and the maximum rate   appears at the SNR $\rho^*$ where the exchange rate of spectral efficiency, quantified by  $\dot{C}(\rho^*)$, ``matches'' the decrease/increase of bandwidth ($\propto 1/\rho^*$).
\end{remark}

Similar observation can also be applied to the mutual information based achievable rate analysis\cite{MedardThesis, MedardTse, Lozano-Porrat, GDME2015} where the spectral efficiency is lower bound by
\begin{align}
 I(X;Y) & = I(X,H;Y) - I(H;Y|X)  \nonumber\\
        & = I(X;Y|H) + I(H;Y) - I(H;Y|X)  \nonumber\\
        &\geq I(X;Y|H) - I(H;Y|X)  \nonumber\\
				&\simeq \log(1+\rho) - \frac{1}{L_c}\log(1+\rho L_c), \label{eqn:MI} 
\end{align}
where the two equalities are from the chain rule, the inequality comes from that fact that $I(H;Y)$ is non-negative, and the approximation is obtained by replacing the two terms by the CSIR/AWGN channel capacity where  $\rho L_c$ is the power boost by using $L_c$ symbols  to estimate one channel coefficient and the pre-log $1/L_c$ is due to the fact that the channel estimation cost is shared by all $L_c$ symbols. For low SNR $\rho\ll 1$,  we can write \eqref{eqn:MI} as 
 \begin{align}
\log(1+\rho) - \frac{1}{L_c}\log(1+\rho L_c) & \simeq \rho-\frac{\rho^2}{2} -(\rho - \frac{\rho^2L_c}{2}) \nonumber\\
& =\frac{\rho^2(L_c-1)}{2},
\end{align}
and for high SNR $\rho\gg 1$ the spectrum efficiency is approximately $\log(1+\rho)$. Therefore, there is also a unique SNR that maximizes the mutual information lower bound \eqref{eqn:MI}.
The exact value of the optimal SNR is determined by the channel coherence length $L_c$ and the corresponding transmission schemes, as shown in Table~\ref{tab:rate_comp}.

To verify the accuracy of the closed-form approximations developed in Sec.~\ref{sec:approx},
in Fig.~\ref{fig:Opt_rho_alpha_Lc_PrN0_20dB_Bc10MHz} we compare the throughput maximizing SNR $\rho^*$ and pilot ratio $\alpha^*$  obtained from the numerical optimization in Sec.~\ref{sec:relax_num} against the results obtained by exhaustive search using \eqref{eqn:Opt_orig}, the 1st-order approximation \eqref{eqn:rho_opt},  and the fine-tuned approximation \eqref{eqn:rho_heur} proposed in Appendix~\ref{app:opt_approximation}. 
From Fig.~\ref{fig:Opt_rho_alpha_Lc_PrN0_20dB_Bc10MHz} we can see that, when the throughput maximizing bandwidth is much larger than the coherence bandwidth, i.e., $\frac{P_r}{N_0}\sqrt[3]{L_c}\gg B_c$, the  discrete-to-continuous relaxation is accurate and there is a good match among the exhaustive search, the numerical optimization, and the closed-form approximation. Even the 1st order approximation  \eqref{eqn:rho_opt} only introduces marginal error.

	\begin{figure}
		\centering
			\includegraphics[width=0.99\columnwidth]{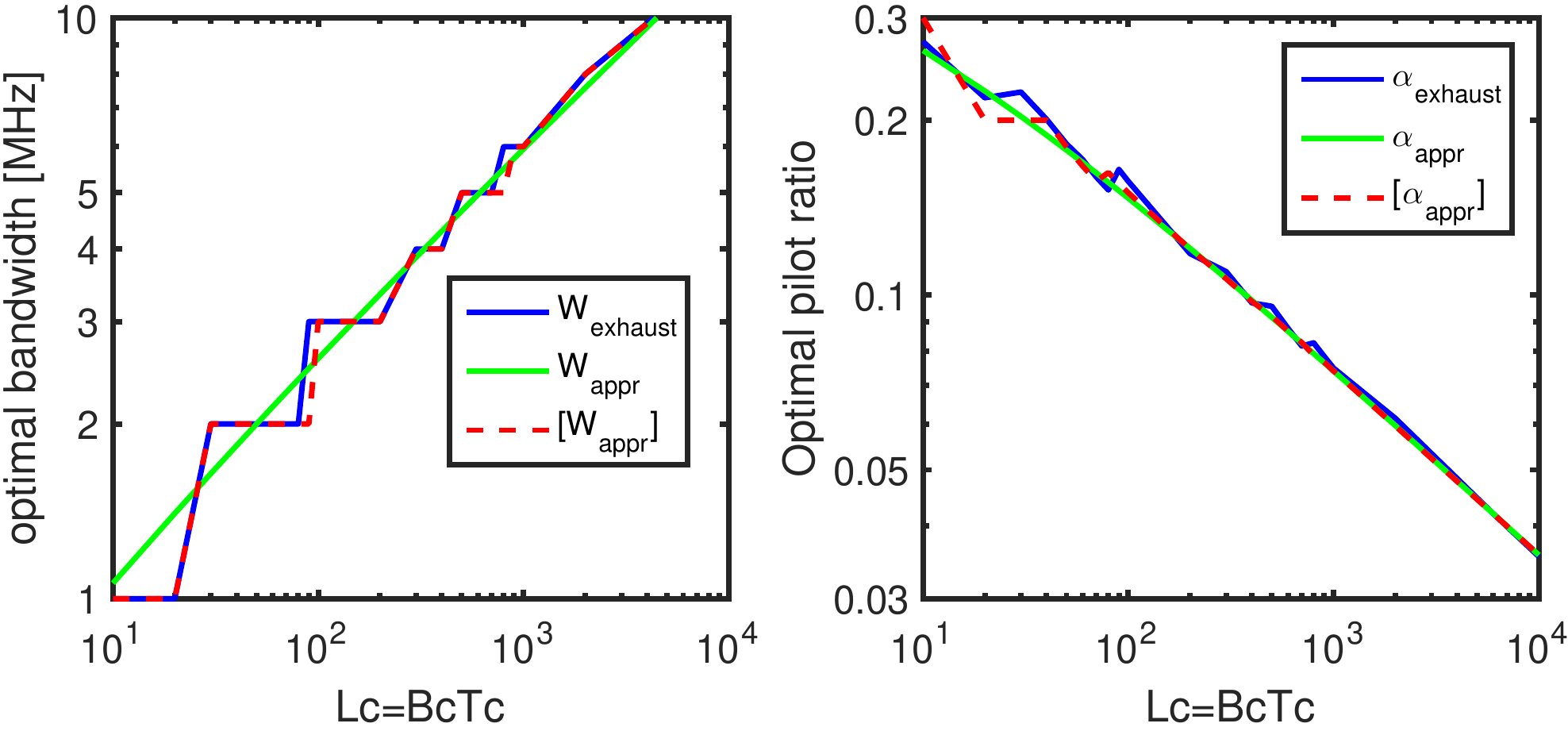}
		\caption{The deviation of the closed-form approximation  from the exhaustive search result 
where the coherence bandwidth $B_c=1$ MHz, coherence time $T_c\in[0.01, 10]$ ms,  and $P_r/N_0=0$ dB.MHz. The mismatch comes from the continuous-value relaxation of the intrinsically discrete valued optimization, which is notable in the parameters selected for this illustration. A round operation of the closed-form approximation to the nearest integer multiplier of coherence $B_c$, denoted by $[W^*]$,  yields  marginal loss. }
		\label{fig:Opt_W_alpha_Lc_PrN0_0dB_Bc1MHz}
	\end{figure}

{Note that $W^*$ and $\alpha^*$ obtained from the numerical optimization will be mapped onto the closest discrete point by rounding.  Since the rounding operation, denoted by $\alpha= [\alpha^*]$, is to ensure that the number of pilots $\alpha L_c$ is an integer, the associated rounding error for $\alpha^*$ is less than $0.5/L_c$ in absolute value or $(2L_c)^{-2/3}$ percentage wise calculated based on \eqref{eqn:alpha_opt}. The rounding operation for $W^*$ to the nearest integer multiplier of coherence $B_c$, denoted by $W=[W^*]$, 
will incur a rounding error for $W^*$ of $B_c/2$ or less in absolute value or $(2L_c)^{-1/3}(\frac{P_r}{N_0B_c})^{-1}$ percentage wise calculated based on \eqref{eqn:rho_opt}. To put them into perspective, for coherence time of $T_c{=}5$ ms and  coherence bandwidth of $B_c{=}10$ MHz (roughly indicates a delay spread of 50--100 ns as reported in mmWave channel models~\cite{3GPP-above6G}), we have $L_c=50000$ and the percentage-wise rounding error is less than $0.06\%$ for $\alpha^*$ and less than $0.3\%$ for $W^*$ with $P_r/(N_0B_c)=10$ dB.} 
On the other hand, when the throughput maximizing bandwidth is comparable with the coherence bandwidth, i.e., $\frac{P_r}{N_0}\sqrt[3]{L_c}=O(B_c)$, the discrete-to-continuous relaxation will introduce some mismatch, as illustrated in Fig.~\ref{fig:Opt_W_alpha_Lc_PrN0_0dB_Bc1MHz} where $\frac{P_r}{N_0}\sqrt[3]{L_c}\in[3B_c, 30B_c]$. Simulation results in Fig.~\ref{fig:Opt_W_alpha_Lc_PrN0_0dB_Bc1MHz} suggest that a round operation  will serve its purpose with marginal loss. For throughput maximizing bandwidth larger than tens of coherence bandwidth, the deviation is hardly visible.

\subsection{Negligible Gain in Optimizing Power for Pilots If Bandwidth is Sufficient}\label{sec:pilot_power}

The rate can be further improved if we also allow different transmit power for pilots and for data symbols, and optimize both the fraction of the pilots and their power level, subject to the same average power constraint. It has been shown in~\cite{Hassibi-Hochwald} that the gain is maximized when we use a single pilot for each channel coherence block (accommodating $L_c$ symbols) with sufficiently high power. 
To quantify the potential gain of optimizing the power of pilots, we use the same parameters from Sec.~\ref{sec:estimation} to represent the power level adjustment for pilots.  Let $\rho_{pilot}=\alpha L_c \rho$ be the SNR of the pilot symbol. Since the SNR for the coherence block of $L_c$ symbols is $\rho$, the SNR for the $(L_c-1)$ data symbols is therefore $\rho_{data}=(1-\alpha)\rho L_c/(L_c-1)$. It is easy to verify that the same average power constraint is respected since 
\[\frac{1}{L_c}\rho_{pilot} + \frac{L_c-1}{L_c}\rho_{data} =  \frac{\alpha L_c \rho}{L_c} + \frac{(L_c-1)}{L_c}\frac{(1-\alpha)\rho L_c}{(L_c-1)} = \rho.\]  
We now substitute $\rho_{pilot}$ into \eqref{eqn:h-estimate} to quantify the channel estimation error and substitute   $\rho_{data}$ into  \eqref{eqn:SNReff} to derive the effective SNR, which leads to the achievable spectral efficiency as
\begin{align}
\frac{R}{W}& = (1-\frac{1}{L_c})\E\left[\log(1+ \frac{\rho_{data}\frac{\rho_{pilot}}{1+\rho_{pilot}}|h|^2}{1+\rho_{data}\frac{1}{1+\rho_{pilot}}})\right]  \nonumber\\ 
& \simeq \E\left[\log(1+ \frac{\alpha(1-\alpha)L_c\rho^2|h|^2}{1+\rho+\alpha L_c\rho -2\alpha\rho})\right], \label{eqn:R_pilot}
\end{align}
where the ``$\simeq$'' in the last step comes from the fact that $(L_c-1)/L_c\simeq 1$ for $L_c\gg1$.

 	If we have sufficient bandwidth, the throughput maximizing SNR $\rho$ is small, and we can approximate \eqref{eqn:R_pilot} by
\begin{align}
\frac{R}{W} {=}  \E\left[\frac{\alpha(1-\alpha)L_c\rho^2|h|^2}{1+\rho+\alpha L_c\rho} \right]\log_2(e) {=} \frac{\alpha(1-\alpha)L_c\rho^2 }{1+\rho+\alpha L_c\rho} \log_2(e),
\end{align}	
which provides the same rate as \eqref{eqn:Rwa} where only the fraction of pilots is optimized. 
Therefore, if bandwidth is not the limiting factor, the gain by also optimizing the power of pilots is negligible. 
However, if bandwidth is limited and hence SNR is high, we have 
\[\log(1+(1-\alpha) \rho_{\mbox{eff}}) > (1-\alpha)\log(1+\rho_{\mbox{eff}}),\] 
and allocating more power to pilot does improve the rate.

\section{Extension to Multiple Antennas}\label{sec:Opt-mimo}

\subsection{SIMO: Antenna Array only at the Receiver}\label{sec:simo}
The transmitter is equipped with a single antenna and the receiver has $N_r$ antennas. Assuming the same number of RF chains as the number of antennas, all the channel coefficients can be estimated by the receiver simultaneously based on the same pilots. Then the estimated channel coefficients are used to perform receive beamforming and the effective SNR of the data symbols in the channel capacity lower bound \eqref{eqn:Rwa} 
can now be written as
\begin{align}
\rho_{\mbox{eff}} =\frac{G\rho \E[|\hat{h}|^2]}{1 + G\rho\E[|\tilde{h}|^2]} 
=  \frac{G\alpha L_c\rho^2}{1 + (G+\alpha L_c)\rho}, \label{eqn:SNReff_SIMO}
\end{align}
where  $\E[|\hat{h}|^2]$  and $\E[|\tilde{h}|^2]$, the power of MMSE channel estimation and its estimation error, respectively, are defined in \eqref{eqn:h-estimate}, and  $G\leq N_r$ is the  gain of signal combining using the estimated channel coefficients.

Following the same procedure as in Appendix~\ref{app:opt_relax}, we can obtain the condition 
\begin{align}
\alpha^2L_c\rho + G\rho(2\alpha-1) = 1-3\alpha. \label{eqn:partial_SIMO} 
\end{align}
Note that by variable substitution $\tilde{\rho}=G\rho$ and $\tilde{L_c}=L_c/G$, we can rewrite \eqref{eqn:SNReff_SIMO} and \eqref{eqn:partial_SIMO}  as
\begin{equation}\label{eqn:rho_SIMO} 
\begin{aligned}
& \rho_{\mbox{eff}} =   \frac{ \alpha \tilde{L_c}\tilde{\rho}^2}{1 + (1+\alpha \tilde{L_c})\tilde{\rho}}, \\
& \alpha^2\tilde{L_c}\tilde{\rho} + \tilde{\rho}(2\alpha-1) = 1-3\alpha. 
\end{aligned}
\end{equation}
Now we can use \eqref{eqn:rho_SIMO} to determine the throughput maximizing $(\tilde{\rho}^*, \tilde{\alpha}^*)$  numerically following the methods developed in Sec.~\ref{sec:relax_num}, and then restore the throughput maximizing $(\rho^*,\alpha^*)$ by
\begin{align}
\rho^* = \tilde{\rho}^*/G, \ \alpha^*= \tilde{\alpha}^*. \label{eqn:optrho_SIMO} 
\end{align}
Similarly,  the closed-form expressions \eqref{eqn:rho_opt}-\eqref{eqn:rate_opt} can now be written as
\begin{align}
&\rho^* \simeq (\frac{4}{L_cG^2})^{1/3}, \ 
\alpha^* \simeq (\frac{G}{2L_c})^{1/3}, \nonumber\\  
&R(\rho^*,\alpha^*) \simeq \left(1-(\frac{4G}{L_c})^{1/3} \right)\frac{P_rG}{N_0}\log_2(e). \label{eqn:approx_SIMO}
\end{align}

\subsection{MISO: Antenna Array only at the Transmitter}\label{sec:miso}

The transmitter is equipped with $N_t$ antennas and the receiver has a single antenna. A limited feedback beam-switching transmission strategy proposed in \cite{CVV2014} is adopted, where the transmitter first send out pilots using a pre-selected beamforming vector out of a pool of $K_t$ vectors. The receiver, after estimating all the superposed channels, one for each beam vector, feeds back to the transmitter the index of 
the ``hottest'' beam. Then the transmitter sends its data symbols using the beamforming vector selected by the receiver.

To simplify our analysis, we assume that the feedback delay (assumed to be much smaller than the channel coherence time $T_c$) and the feedback data rate (used to transmit the index of the beamforming vector, only a few bits) is negligible. Then on average each beam is estimated using $\alpha L_c/K_t$ pilots, and the MMSE channel estimation \eqref{eqn:h-estimate} of the selected beam (will be used for data transmission) can be written as
\begin{align}
 \E[|\hat{h}|^2]=\frac{ G\rho\alpha L_c/K_t}{1+ G\rho\alpha L_c/K_t}, \ \E[|\tilde{h}|^2]=\frac{1}{1+G\rho\alpha L_c/K_t }, \label{eqn:hest_MISO}
\end{align}
where $G\leq N_t$ is the average gain of the hottest beam selected. The corresponding effective SNR \eqref{eqn:SNReff} can then be written as 
\begin{align}
\rho_{\mbox{eff}} =\frac{G\rho \E[|\hat{h}|^2]}{1 + G\rho\E[|\tilde{h}|^2]} 
=  \frac{(G\rho)^2\alpha L_c/K_t}{1 + (1+\alpha L_c/K_t)G\rho}. \label{eqn:SNReff_MISO}
\end{align}

Following the same procedure as in Appendix~\ref{app:opt_relax}, we can obtain the condition 
\begin{align}
\alpha^2\frac{L_c}{K_t}G\rho + G\rho(2\alpha-1) = 1-3\alpha. \label{eqn:partial_MISO} 
\end{align}
As in Sec.~\ref{sec:simo}, we  now use  variable substitution $\tilde{\rho}=G\rho$ and $\tilde{L_c}=L_c/K_t$  to obtain  \eqref{eqn:rho_SIMO}, which can be used to determine 
the throughput maximizing $(\tilde{\rho}^*, \tilde{\alpha}^*)$ numerically. Note that we have $\rho^*=\tilde{\rho}^*/G$ and $\alpha^*=\tilde{\alpha}^*$. Similarly, we can obtain the closed-form approximations as follows
\begin{align}
&\rho^* \simeq \frac{1}{G}(\frac{4K_t}{L_c})^{1/3}, \ 
\alpha^* \simeq (\frac{K_t}{2L_c})^{1/3}, \nonumber\\  
&R(\rho^*,\alpha^*) \simeq \left(1-(\frac{4K_t}{L_c})^{1/3} \right)\frac{P_rG}{N_0}\log_2(e). \label{eqn:approx_MISO}
\end{align}

\subsection{MIMO: Antenna Arrays at both the Transmitter and the Receiver}\label{sec:mimo}

The transmitter is equipped with $N_t$ antennas and the receiver has $N_r$ antennas. When there is no constraint on bandwidth, the throughput maximizing transmission strategy is to use as much bandwidth as needed to maximize the throughput, which necessarily put the overall transmission into the low SNR regime~\cite{Lozano-Porrat, GDME2015}, as also witnessed in Sec.~\ref{sec:simo} and Sec.~\ref{sec:miso} where the throughput maximizing operation SNR $G\rho^*$, with beamforming gain,  turns out to be small. In the low SNR regime, rank-1 transmission strategy is near optimal since  
\[\E[\log(|I + \rho H^{\dag}H|)] = \E[\log(1 + \rho\trace(H^{\dag}H))] + O(\rho^2),\ \rho<1.\]

As in Sec.~\ref{sec:miso} we focus on the beam-switching transmission strategy~\cite{CVV2014}, where the transmitter maintains a pool of $K_t$ pre-selected beamforming vectors and sweeps over all the candidate beams by transmitting dedicated pilots. The receiver, after estimating all the superposed channels, $N_t$ of them per each receiver antenna, feeds back to the transmitter the index of the ``hottest'' beam that provides the highest  gain after applying signal combining at the receiver. Then the transmitter sends its data symbols using the beamforming vector selected by the receiver and the receiver applies the corresponding best signal combining vector for the selected beam using the estimated channel coefficients. 

Let $G_1\leq N_t$ be the expectation of the gain of the ``hottest'' beam  selected by the beam-switch mechanism, the MMSE channel estimation of the best beam can be characterized, following \eqref{eqn:hest_MISO}, as
\begin{align}
 \E[|\hat{h}|^2]=\frac{ G_1\rho\alpha L_c/K_t}{1+ G_1\rho\alpha L_c/K_t}, \ \E[|\tilde{h}|^2]=\frac{1}{1+G_1\rho\alpha L_c/K_t }. \label{eqn:hest_MIMO}
\end{align}
Let $G_2\leq N_r$ denote the gain of receive signal combining, the combined transmit-receive beamforming gain would be $G_1G_2$ and the corresponding effective SNR of the data symbols can be written as 
	\begin{align}
\rho_{\mbox{eff}} & =\frac{G_1G_2\rho \E[|\hat{h}|^2]}{1 + G_1G_2\rho\E[|\tilde{h}|^2]} 
=  \frac{G_2(G_1\rho)^2\alpha L_c/K_t}{1 + (G_2+\alpha L_c/K_t)G_1\rho} \nonumber\\
& = \frac{(G_1G_2\rho)^2\alpha \frac{L_c}{K_tG_2}}{1 + (1 +\alpha \frac{L_c}{K_tG_2})G_1G_2\rho}. \label{eqn:SNReff_MIMO}
\end{align}
 
\begin{remark}
Note that the gain $G_1$, which is obtained by selecting the ``hottest'' beam out of $K_t$ candidate beams using $N_t$ transmit antennas, is materialized both during  channel estimation  and  data transmission process, as shown in \eqref{eqn:hest_MIMO} and \eqref{eqn:SNReff_MIMO}. The  gain $G_2$, on other hand, is obtained by performing receive signal combining based on the estimated channel coefficients. Therefore $G_2$ is only available during the data transmission stage rather than the channel estimation stage, as shown in \eqref{eqn:hest_MIMO}.
\end{remark}
	
Following similar steps as in Sec.~\ref{sec:simo} and  Sec.~\ref{sec:miso}, we  now use  variable substitution $\tilde{\rho}=G_1G_2\rho$ and $\tilde{L_c}=L_c/(K_tG_2)$  to obtain  \eqref{eqn:rho_SIMO}, which can then be used to determine 
the throughput maximizing $(\tilde{\rho}^*, \tilde{\alpha}^*)$ numerically, and the closed-form approximations as follows
\begin{align}
& \rho^* \simeq \frac{1}{G_1G_2}(\frac{4K_tG_2}{L_c})^{1/3}, \ 
\alpha^* \simeq (\frac{K_tG_2}{2L_c})^{1/3}, \nonumber\\    
& R(\rho^*,\alpha^*) \simeq \left(1-(\frac{4K_tG_2}{L_c})^{1/3} \right)\frac{P_rG_1G_2}{N_0}\log_2(e). \label{eqn:approx_MIMO}
\end{align}	

\begin{remark}
\eqref{eqn:approx_MIMO} and its associated variable substitution $\tilde{\rho}=G_1G_2\rho$ and $\tilde{L_c}=L_c/(K_tG_2)$ are generic. We can recover  the SIMO case by setting $K_t=G_1=1$ and recover   the MISO case with beam-switching by setting $G_2=1$. If beam-switching is not used and the full channel matrix is estimated before choosing the beamforming vectors for data transmission, we can set $K_t=N_t$ and $G_1=1$, and leave $G_2$ to represent the actual  gain during the data transmission process. On the other hand, if beam-switching strategy is adopted both at the transmitter and the receiver, we can use $K_t\sim O(N_tN_r)$ to represent the size of overall  candidate beams and let $G_1\leq N_tN_r$ represent the gain of the ``hottest'' beam, leaving $G_2=1$  to reflect the fact that there is no extra combining gain in the data transmission stage as compared to the channel estimation stage.  
\end{remark}

\begin{figure*} 
	\centering
		\includegraphics[width=\textwidth]{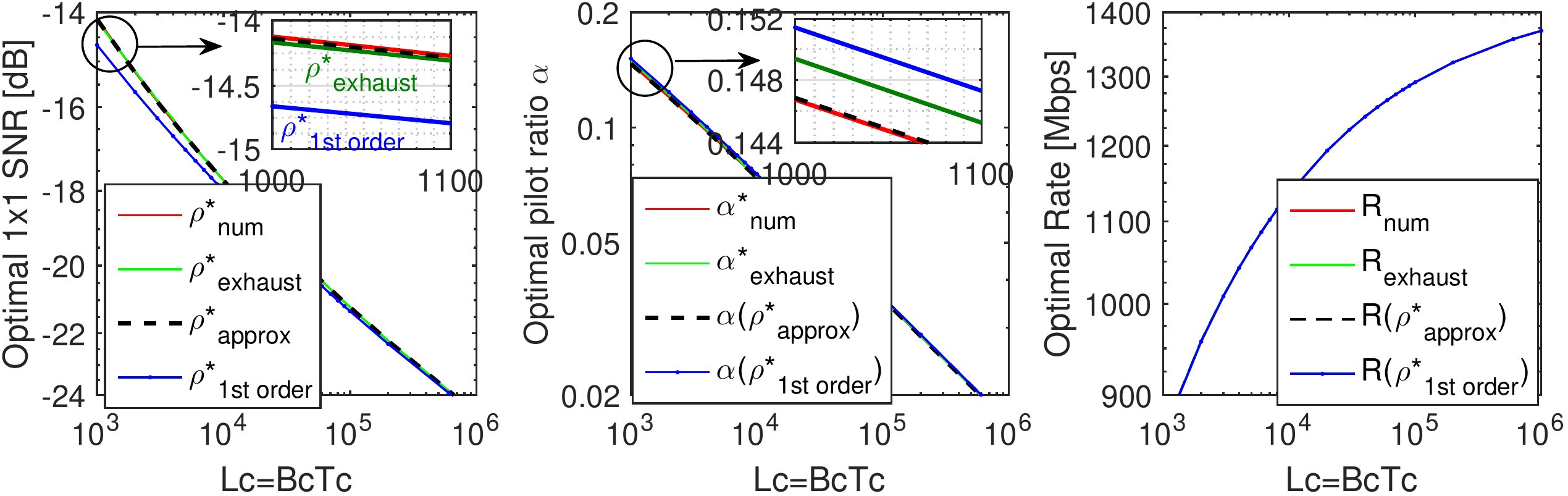}
	\caption{Comparison of the throughput maximizing SNR $\rho^*$ obtained by exhaustive search for \eqref{eqn:Opt_orig}, numerical optimization using \eqref{eqn:partial_MISO}, the 1st-order approximation \eqref{eqn:MIMO_opt}, and the closed-form approximation by resorting to \eqref{eqn:rho_heur}. The throughput maximizing pilot ratio $\alpha^*$ is obtained using \eqref{eqn:partial_MISO} and the corresponding achievable rates are achieved using corresponding $(\rho^*,\alpha^*)$.   Coherence bandwidth $B_c=10$ MHz, coherence time $T_c\in[0.1, 100]$ ms,  and $P_r/N_0=20$ dB.MHz, and beam-switching with full gain $N_tN_r=10$ are used for the simulation.}
	\label{fig:Opt_rho_alpha_Lc_PrN0_20dB_Bc10MHz_MIMO}
\end{figure*}

In Fig.~\ref{fig:Opt_rho_alpha_Lc_PrN0_20dB_Bc10MHz_MIMO} we evaluate the accuracy of the closed-form approximations of the throughput maximizing SNR $\rho^*$ (hence $W^*$) for MIMO setup, where beam-switching is adopted assuming full combining gain $N_tN_r=10$. Compared with the results obtained from numerical optimization and from exhaustive search, the 1st-order closed-form approximation of $\rho^*$ in \eqref{eqn:MIMO_opt} admits some small error when the channel coherence $L_c$ is not sufficiently large, and the error diminishes as $L_c\to\infty$. The associated error on pilot ratio and achievable rate, caused by inaccurate $\rho^*$ approximation, is negligible. The results obtained by resorting to the closed-form approximation \eqref{eqn:rho_heur} provide excellent match with the numerical optimization and the exhaustive search.

\subsection{Impact of Beam-Switching Strategy and the Hottest Beam}\label{sec:BF-BS}

It is worthwhile to examine the operational parameters $(\rho^*, \alpha^*, R(\rho^*,\alpha^*))$ for some special cases to see the impact of the beam-switching strategy.
Assuming beam-switching is used both at the transmitter side with $N_t$ candidate beams and at the receiver side with $N_r$ candidate beams, the expected gain of the ``hottest''	beam,  selected by sweeping over all candidate beams, depends on the number of antennas $N_t$ and $N_r$, pre-selected beamforming vector codebook, and channel characteristics.

For example,  when channel is highly directional (e.g., small angular spread) and the beamforming vector codebook is designed based on a prior knowledge of the channel learned either from previous feedback or from the approximate uplink/downlink channel reciprocity, the overall beamforming/combining gain for the hottest beam is close to $N_tN_r$. By substituting $G_1=N_rN_t$, $K_t=N_tN_r$, and $G_2=1$ into \eqref{eqn:approx_MIMO}, we obtain
\begin{align}
&\rho^* \simeq \frac{1}{N_tN_r}(\frac{4N_tN_r}{L_c})^{1/3}, \ 
\alpha^* \simeq (\frac{N_tN_r}{2L_c})^{1/3}, \nonumber\\  
& R(\rho^*,\alpha^*) \simeq \left(1-(\frac{4N_tN_r}{L_c})^{1/3} \right)\frac{P_rN_tN_r}{N_0}\log_2(e). \label{eqn:MIMO_opt}
\end{align}

However, if the channel has rich scattering and all the entries in the full-rank channel matrix are I.I.D. with finite variance, the average gain of the ``hottest''	beam can be approximated\footnote{Strictly speaking, the gain of the ``hottest'' beam is smaller than the gain of the strongest channel eigenmode when we do not have the full CSI for signal combining. Here we approximate the gain of the ``hottest'' beam by the largest eigenvalue $\lambda_{\mbox{max}}(\bH^*\bH)$, whose distribution can be found in~\cite{Karoui2006,TracyWidom}.} as $O(N_t+N_r)$ instead of $N_tN_r$. By setting $G_1=N_t{+}N_r$ we will have
\begin{align}
& \rho^* \simeq \frac{1}{N_t{+}N_r}(\frac{4N_tN_r}{L_c})^{1/3}, \ 
\alpha^* \simeq (\frac{N_tN_r}{2L_c})^{1/3}, \nonumber\\    
& R(\rho^*,\alpha^*) \simeq \left(1-(\frac{4N_tN_r}{L_c})^{1/3} \right)\frac{P_r(N_t{+}N_r)}{N_0}\log_2(e). \label{eqn:MIMO_pessi2}
\end{align}

\section{How Much Bandwidth is too Much: Numerical Results}\label{sec:num}

In  this section  we evaluate a  $N_t\times N_r$  MIMO link {in mmWave band} where the transmitter sweeps  for the hottest beam out of $N_t$ transmit candidate beams and the receiver sweeps over $N_r$ candidate beams.  Carrier frequency $f_c$, path loss, transmission distance $d$, and transmit power $P_t$ are subject to design.  Unless otherwise specified, in this section we assume $T_c\times B_c$ block fading channels with coherence time $T_c=5$ ms and coherence bandwidth $B_c=10$ MHz\footnote{{Coherence time of $T_c{=}5$ ms roughly matches user/environment movement at speed of 1 m/s at 28 GHz, and the coherence bandwidth of $B_c{=}10$ MHz roughly indicates a delay spread of 50--100 ns as suggested by mmWave channel measurements~\cite{3GPP-above6G}.}}, resulting in a channel coherence length $L_c=5\times 10^4$. 
Each transmit antenna has element gain $G_t=8$ dBi and each receive antenna has element gain  $G_r=5$ dBi. The noise figure is set to $F=9$ dB.
 
\subsection{Maximum Rate and Optimal Bandwidth for 28 GHz Band at 100-Meter Distance}

\begin{figure*} 
	\centering 
		\includegraphics[width=0.89\textwidth]{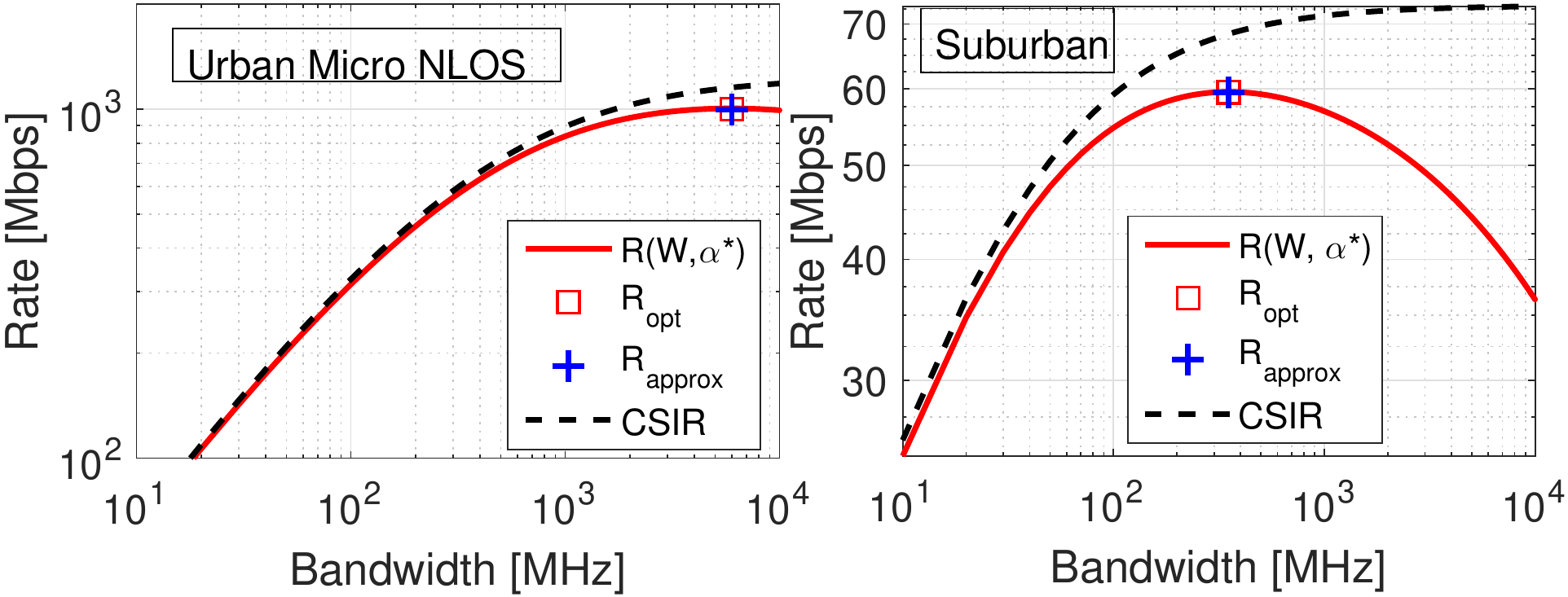} 
	\caption{The maximized rate at transmission distance $d=100$ meters as a function of signaling bandwidth for a 28GHz band link using $P_t=1$ Watt transmit power with $N_t=16$  antenna elements at the transmitter and $N_r=2$ elements at the receiver: (left) 3GPP Urban Micro NLOS~\cite{3GPP-below6G}; (right) 5GCM Suburban~\cite{5GCM-VzW}. Ideal beamforming with rank-1 transmission (i.e., single data stream) is assumed. The rate assuming perfect CSIR, the maximum	$R(W^*,\alpha^*)$ and the throughput maximizing bandwidth $W^*$, and their closed-form approximation counter parts obtained using \eqref{eqn:rho_heur} and \eqref{eqn:alpha_heur} are plotted as references.}
	\label{fig:Opt_rate_W_1Watt_28GHz}
\end{figure*}

In Fig.~\ref{fig:Opt_rate_W_1Watt_28GHz} we plot the maximum rate as a function of signaling bandwidth for a link at the $f_c=28$ GHz band using  $P_t=1$ Watt transmit power with $N_t=16$  antenna elements at the transmitter and $N_r=2$ elements at the receiver, placed at 100 meters apart. Ideal beamforming with rank-1 transmission (i.e., single data stream) is assumed and the percentage of pilots is optimized to maximize the rate. When channel coefficients are known at the receiver (CSIR), there is no channel estimation penalty and therefore using more bandwidth will monotonically increase the achievable rate. Under the 3GPP Urban Micro NLOS~\cite{3GPP-below6G} path loss model\footnote{Note that \cite{3GPP-below6G} is designed for frequency below 6GHz and we simply apply frequency extrapolation to high bands, since in \cite{3GPP-above6G} the Urban Micro is only defined for Street Canyon scenario.}, the maximum rate peaks at $W^*\cong 6$ GHz, which is far more than the available bandwidth assigned by FCC (two channels each of 425MHz). When the path loss is severe, such as scenarios modeled by the 5GCM Suburban channel model\footnote{{This channel model is designed for the scenario where the base station antennas are placed below clutter top and user equipment is under clutter, with 12 dB vegetation loss to account for wave propagation over trees.}}~\cite{5GCM-VzW}, the maximum rate peaks at $W^*\leq 400$ MHz, and using more bandwidth will hurts the achievable rate. As the  bandwidth approaches its optimum,  adding more bandwidth leads to very limited rate gain: in both cases, using one tenth of the optimal bandwidth $W^*$ only leads to less than $30\%$ rate deduction. Also note that our closed-form approximation matches well with the numerical solutions, which means bandwidth optimization can be performed on-the-fly using our closed-form expressions.

\subsection{Maximum Beneficial Bandwidth as a Function of Transmission Distance}

\begin{figure*} 
	\centering
	(a) $32\times4$ MIMO under Blocked LOS path loss\\ \vspace{3mm}
		\includegraphics[width= \textwidth]{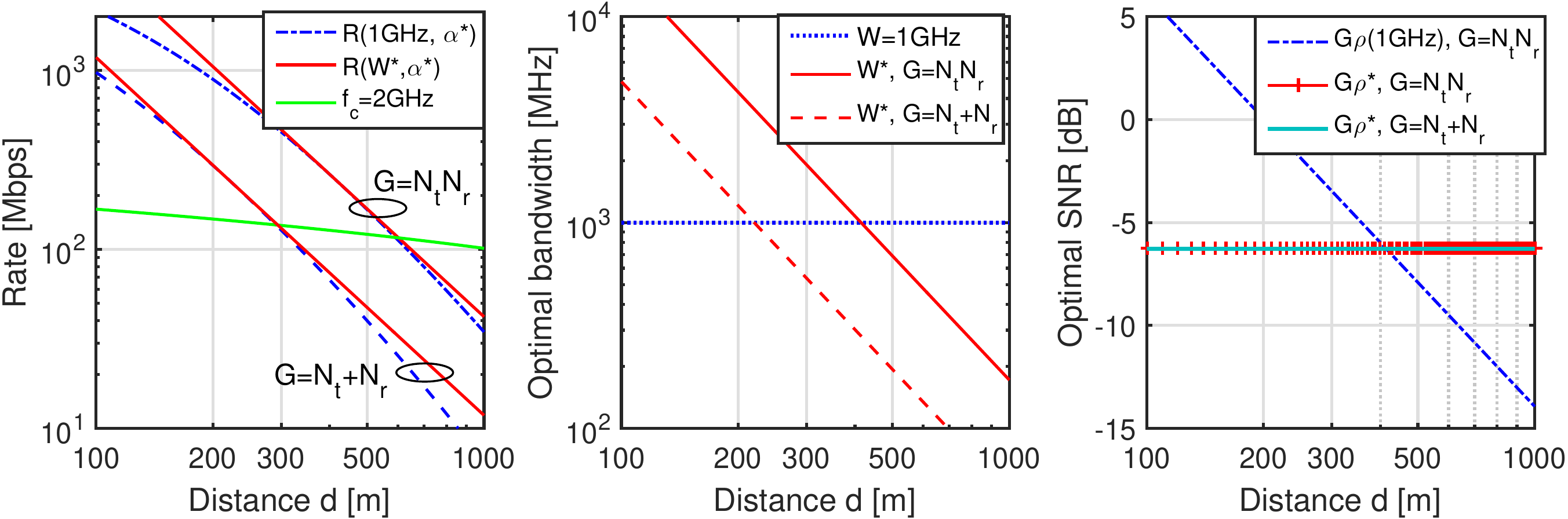}\\
	(b) $64\times4$ MIMO under 3GPP uMi-NLOS path loss\\	\vspace{3mm}
		\includegraphics[width= \textwidth]{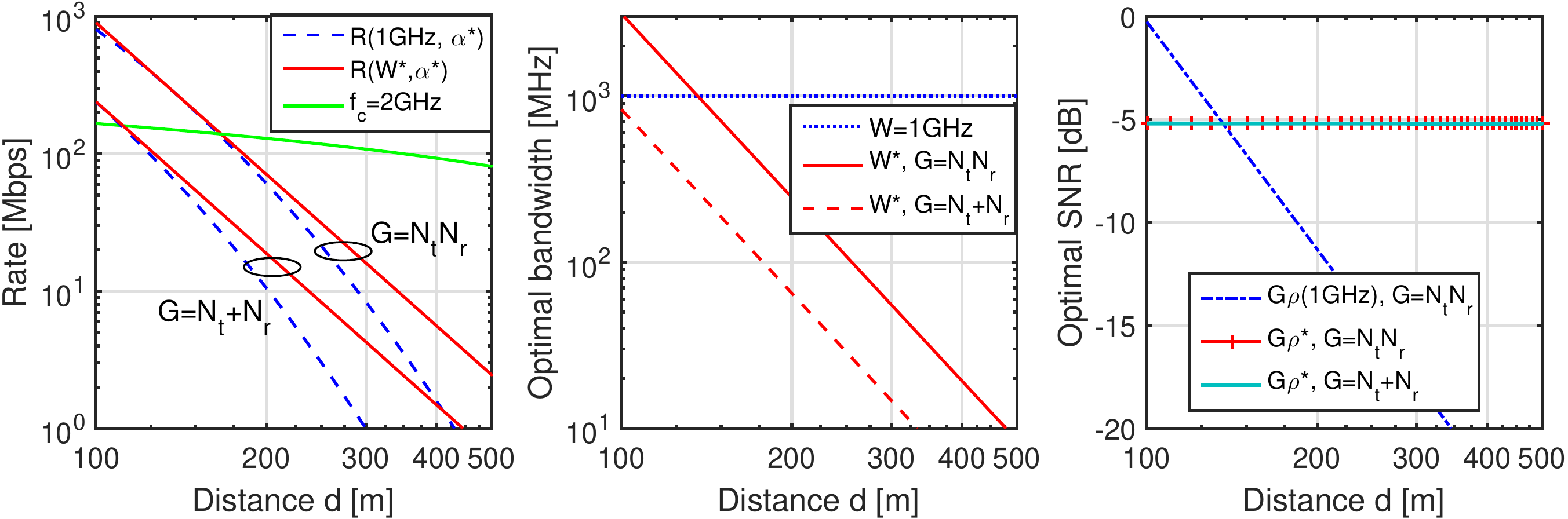} 
	\caption{The maximized Rate $R(W^*,\alpha^*)$, the throughput maximizing bandwidth $W^*$,  and corresponding SNR $G\rho^*$ for a $N_t\times N_r$ MIMO link with $P_t=30$ dBm (1 Watt) at $f_c=60$ GHz band. Joint TX-RX beam-switching is used and the average gain $G$ of the hottest beam is assumed to be either $N_tN_r$ or $N_t{+}N_r$, which represent the ideal case (high directional channel) and the rich scattering case (i.i.d. entries in channel matrix), respectively. The case with fixed bandwidth $W=1$ GHz using optimized pilot ratio $\alpha^*$ are plotted to illustrate the benefit of bandwidth optimization. The low band SISO case with carrier $f_c=2$ GHz, bandwidth $W=10$ MHz, and transmit power $P_t=46$ dBm (40 Watt) is also plotted as a benchmark. (a)  blocked LOS (i.e., free-space path loss with 25dB loss~\cite{CVV2015}), (b) 3GPP uMi-NLOS~\cite{3GPP-below6G}.}
	\label{fig:Opt_rate_W_BLOS_MIMO}
\end{figure*}

In Fig.~\ref{fig:Opt_rate_W_BLOS_MIMO}  we demonstrate the influence of beamforming gain of the hottest beam on 
the maximized Rate $R(W^*,\alpha^*)$, throughput maximizing bandwidth $W^*$,  and corresponding SNR $G\rho^*$ for a $N_t\times N_r$ MIMO link with $P_t=30$ dBm (1 Watt) at $f_c=60$ GHz band. 
The average gain $G$ of the hottest beam is assumed to be either $N_tN_r$ or $N_t{+}N_r$, which represent the ideal case with high directional channels and the rich scattering case, respectively.

When the path loss is favorable, as shown in Fig.~\ref{fig:Opt_rate_W_BLOS_MIMO}  (a) for 
the Blocked LOS path loss model~\cite{CVV2015} (i.e., free-space path loss plus 25dB shadowing loss {to account for blockage effect in mmWave}), even in the ideal  case, the achievable rate is less than 1~Gbps for transmission distance beyond 200 meters.  
The maximum beneficial bandwidth $W^*$ (i.e., the bandwidth that maximizes the rate) decreases with the transmission distance and it becomes less than 200 MHz for distance beyond 1000 meters. 
When the path loss is severe, as shown in Fig.~\ref{fig:Opt_rate_W_BLOS_MIMO}  (b) for 
the 3GPP uMi-NLOS~\cite{3GPP-below6G} path loss, even in the ideal case, the achievable rate is less than 1~Gbps for transmission distance beyond 100 meters, and the rate is below the low-band benchmark (SISO with $f_c{=}2$ GHz, $W{=}10$ MHz, $P_t{=}46$ dBm) for distance beyond 160 meters.  The maximum beneficial bandwidth $W^*$ is less than 1~GHz for distance beyond 140 meters and it becomes less than 100~MHz for range beyond 260 meters.

For the rich scattering case where channel matrix contains i.i.d. entries, with the average gain of the hottest beam at $N_t{+}N_r$, the rate and the throughput maximizing bandwidth are roughly $\min(N_t,N_r)$ times lower as compared to the ideal case, since bandwidth optimization will provide rates almost linearly with the signal power.

\subsection{Maximum Beneficial Bandwidth as a Function of EIRP}

\begin{table*}[t]
	\centering
		\caption{Potential Achievable EIRP for mmWave Band}
	\label{tab:EIRP}
		\begin{tabular}{|c|c|c|c|c|}
		\hline
		Type & No. Elements  & Power per element  & Element gain $G_t$ & EIRP  \\ \hline
		30 dBm sum power &  $N_t$  & --- &  8 dBi  &  $38+10\log_{10}(N_t)$ dBm  \\ \hline
		RF IC \cite{Shahriar2015} & $N_t=16$ & 10 dBm & 2 dBi & 36 dBm \\	\hline
    Hydra chip~\cite{Wright2016} & $N_t=16$  & 20 dBm & 8 dBi & 52 dBm \\ \hline
		large array & $N_t$  & 20 dBm & 8 dBi & $28 + 20\log_{10}(N_t)$ dBm \\ \hline
		FCC regulation \cite{FCC2016} & --- & --- & ---  & $\leq 75$ dBm per 100 MHz\\
		\hline
		\end{tabular}
\end{table*}

\begin{figure*}
	\centering
		(a) $f_c=28$ GHz\\ \vspace{2.5mm}
		\includegraphics[width= \textwidth]{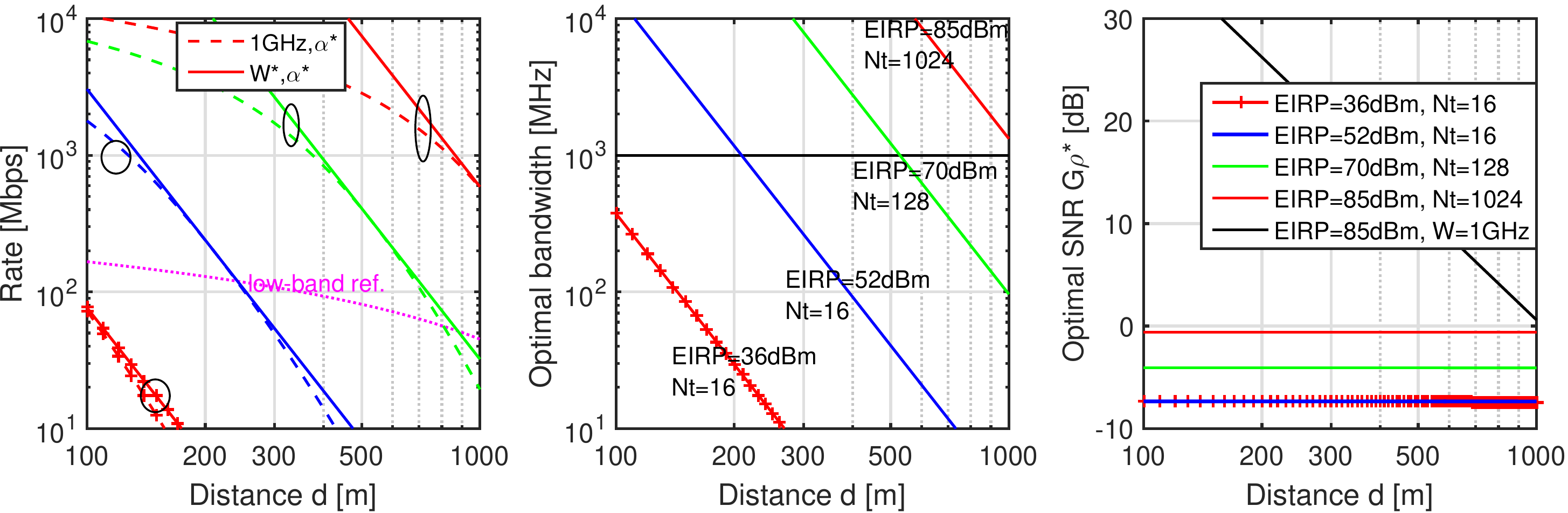}\\
	  (b) $f_c=39$ GHz\\ \vspace{2.5mm}
		\includegraphics[width= \textwidth]{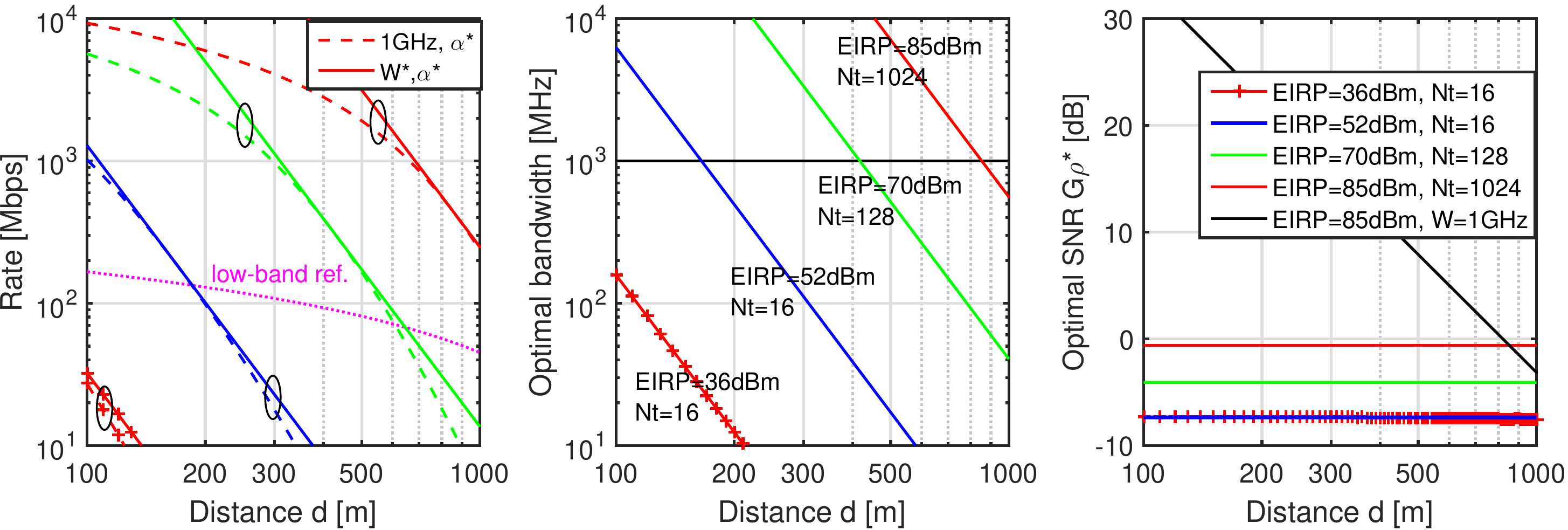}\\
		(c) $f_c=60$ GHz\\ \vspace{2.5mm}
			\includegraphics[width= \textwidth]{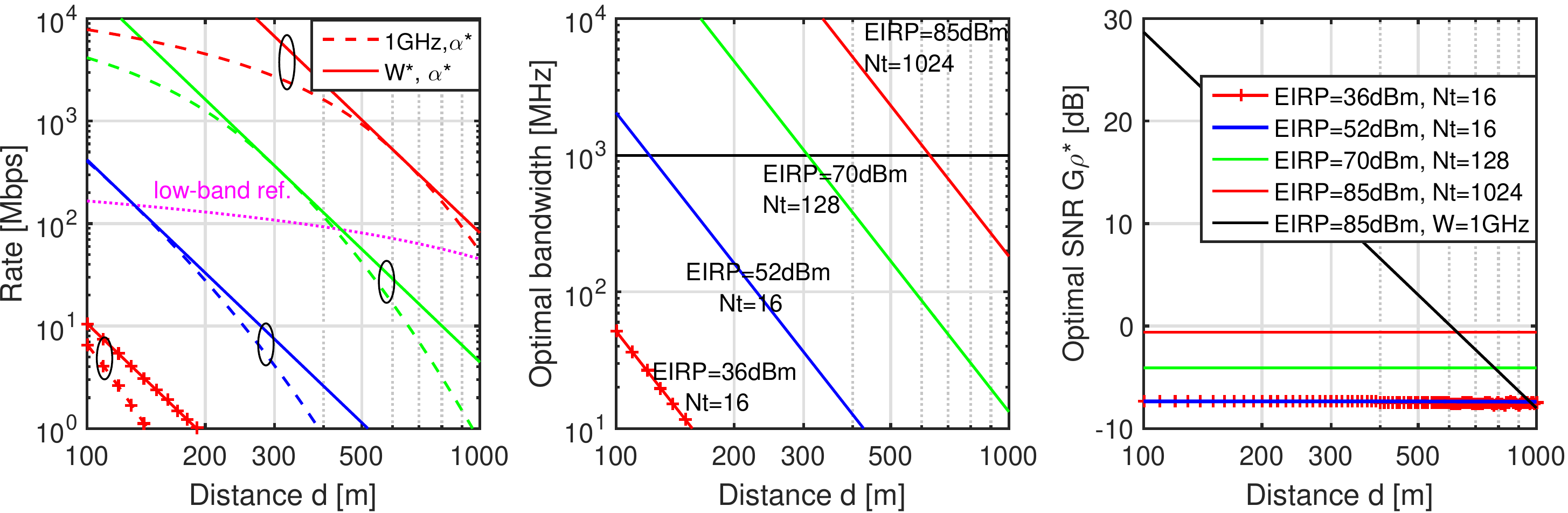} 
	\caption{The maximized Rate $R(W^*,\alpha^*)$, the throughput maximizing bandwidth $W^*$,  and corresponding SNR $G\rho^*$ for a $N_t\times 4$ MIMO link with different EIRP and transmit antennas $N_t$ adopted from Table~\ref{tab:EIRP}. Joint TX-RX beam-switching is used with the average gain of the hottest beam  $G=N_tN_r$ (ideal case), and 3GPP uMi-NLOS path loss model is used. The case with fixed bandwidth $W=1$ GHz (dash lines) using optimized pilot ratio $\alpha^*$ are plotted to illustrate the benefit of bandwidth optimization. The low band SISO case with carrier $f_c=2$ GHz, bandwidth $W=10$ MHz, and transmit power $P_t=46$ dBm (40 Watt) is also plotted (dotted line) as a benchmark. (a) $f_c=28$ GHz band, (b) $f_c=39$ GHz band, (c) $f_c=60$ GHz band.}
	\label{fig:Opt_EIRP_uMi-NLOS_MIMO}
\end{figure*}
 
In Table~\ref{tab:EIRP} we list the potential  EIRP that might be supported by next generation RF techniques. 
The EIRP upper limit of $75$~dBm per 100~MHz, set by FCC regulation~\cite{FCC2016},  is also 
listed for reference.  
In Fig.~\ref{fig:Opt_EIRP_uMi-NLOS_MIMO} we demonstrate the maximized Rate $R(W^*,\alpha^*)$, the throughput maximizing bandwidth $W^*$,  and corresponding SNR $G\rho^*$ for a $N_t\times 4$ MIMO link at (a) $f_c=28$ GHz  band (b) $f_c=39$ GHz  band, and (c) $f_c=60$ GHz  band,  with different EIRP and transmit antennas. Joint TX-RX beam-switching is assumed with the average gain of the hottest beam as $G{=}N_tN_r$ (ideal case), and 3GPP uMi-NLOS path loss model~\cite{3GPP-below6G} is used with frequency extrapolation. 
With 70~dBm EIRP and 1~GHz bandwidth, 1~Gbps user rate at 28GHz/39GHz/60GHz band can be provided  up to 400/310/220  meters.
At the EIRP ceiling set by the FCC regulation, i.e., 85 dBm EIRP using 1 GHz bandwidth,  1~Gbps rate can be delivered up to 860/680/500~meters.

The maximum beneficial bandwidth with 70~dBm EIRP is less than 1~GHz for users beyond 530/420/310~meters, and it becomes less than 100~MHz when the transmission distance is longer than 1000/780/580~meters. 
The EIRP ceiling set by the FCC regulation allows a maximum beneficial bandwidth of 1~GHz bandwidth at range beyond 1080/850/630~meters. On the other hand, at 52~dBm EIRP~\cite{Wright2016}, the maximum beneficial bandwidth is less than 1~GHz for users beyond 210/170/120~meters with corresponding rate at around 200~Mbps. The maximum beneficial bandwidth is less than 100~MHz when the transmission distance is longer than 400/310/230~meters.

\section{Potential Use Cases}\label{sec:use}

 \begin{figure*} 
	\centering
	(a) 28 GHz with $N_t=24$ and $N_r=1$   \hspace{2cm} 
	(b) 39 GHz with $N_t=24$ and $N_r=1$ \\ 	\vspace{3.5mm}  
	\includegraphics[width=0.48\textwidth]{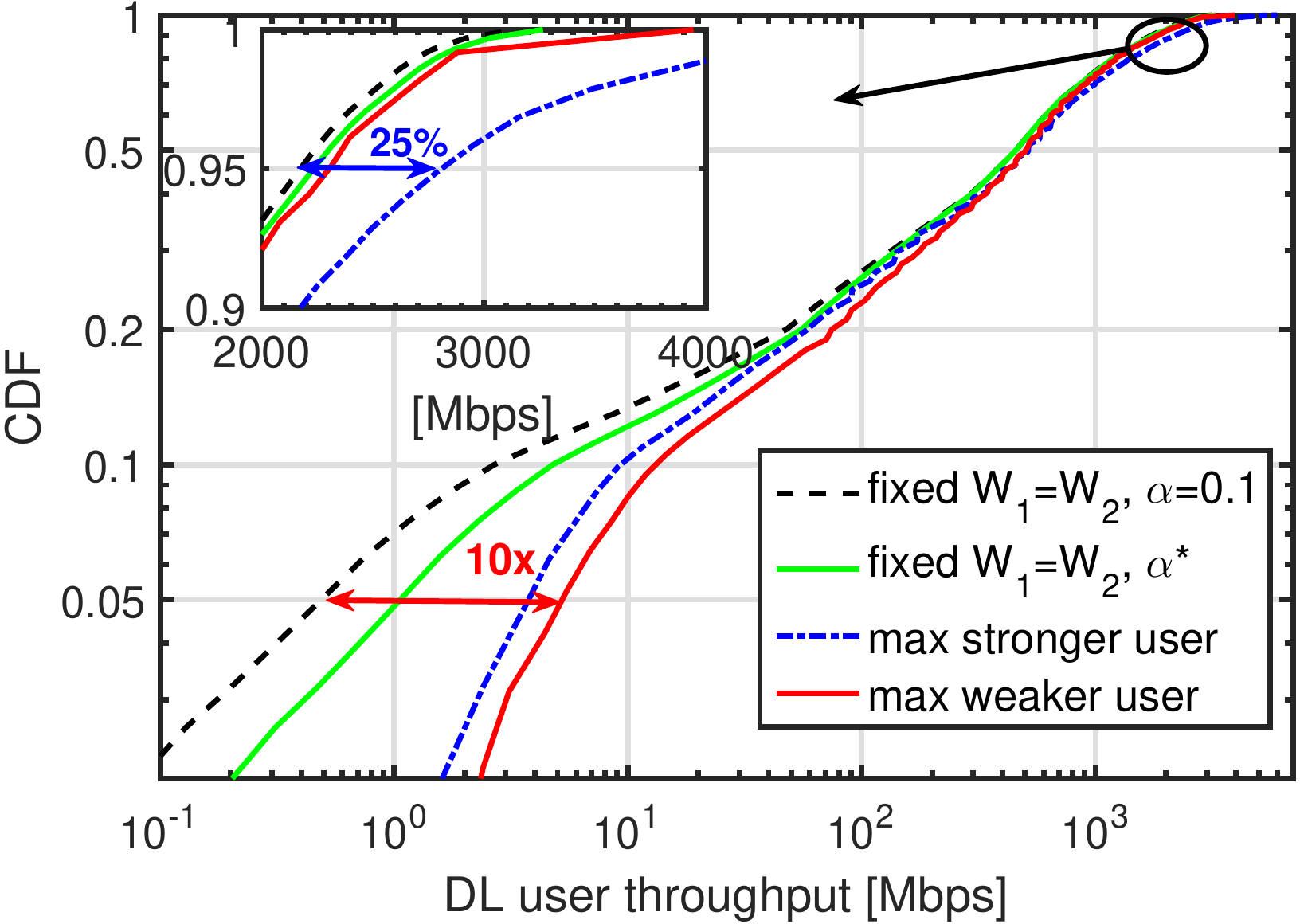} \hspace{1mm}
	\includegraphics[width=0.48\textwidth]{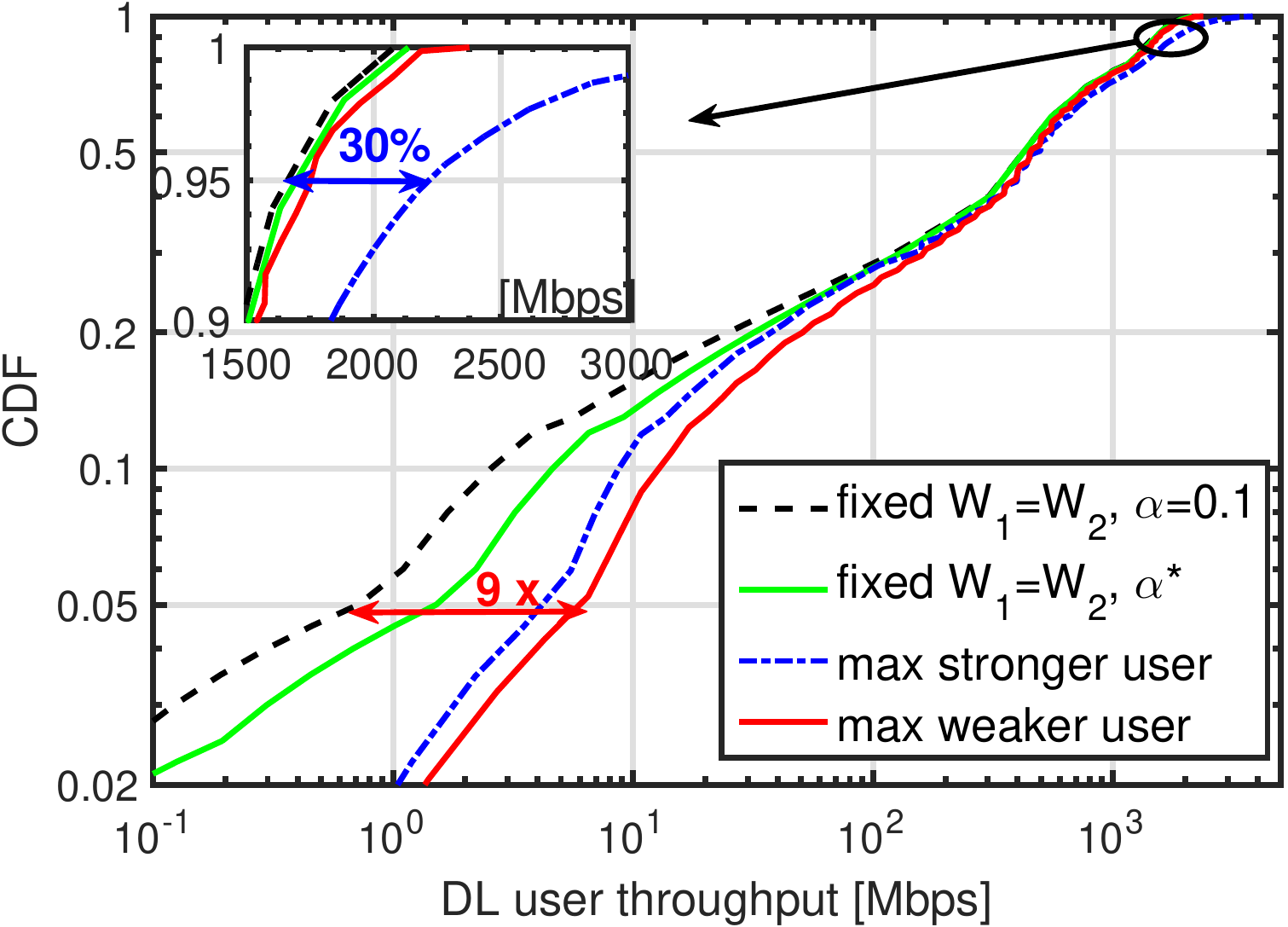} 
	\caption{CDF of downlink throughput by two-user bandwidth-power joint  optimization subject to the constraint that no one's rate shall be decreased after the joint allocation. The  user throughput is evaluated by leveraging the CDFs of SINR generated by~\cite{Ozge2015}  
	for the 28 GHz band and by~\cite{Vook2016}  
	for the 39 GHz band. (a)  28 GHz system with 228 meters ISD equipped with $24\times1$ MIMO with transmit power $P_i=24$dBm and bandwidth $W_i=250$MHz budget for each user; (b)  39 GHz system with 150 meters average ISD equipped with $32\times1$ MIMO subject to  68 dBm EIRP constraint and  $W_i=200$MHz per user bandwidth budget. For convenience, the channel coherence is assumed to be $L_c=2.4\times10^4$ and  $L_c=3.2\times10^4$, respectively. }
	\label{fig:CDF_Rate_BW-Power-pooling}
\end{figure*}

\subsection{New Air Interface: Constant Spectral Efficiency Transmission}

As discussed in  Sec.~\ref{sec:Opt-siso} and  Sec.~\ref{sec:Opt-mimo}, when signaling bandwidth is free to be optimized, the maximized $1\times1$ SNR $\rho^*$ turns out to be a constant that depends on channel coherence $L_c$ and beamforming gain $G$, and their dependence can be approximately described as  
\[\rho^*  \simeq \frac{1}{G}\sqrt[3]{\frac{4G}{L_c}}.\]
When channel coherence $L_c$ and beamforming gain $G$ remain unchanged, the throughput maximizing bandwidth $W^*$ increases as the user becomes closer to the base station and decreases when the user moves away such that the overall SNR $G\rho^*$ as well as the pilot ratio $\alpha^*$ remain constant. Therefore, a constant low spectral efficiency transmission is optimal in the sense that it maximizes the user's rate.  This can be an attractive option {for new air interface in mmWave band} since it will allow the transceiver to use {the same coding/modulation schemes without sacrificing performance, in contrast to the classical adaptive coding and modulation look-up tables used in the current telecommunication systems.  The price to pay is the coordination between the transmitter and the receiver (e.g., via feedback) to adapt the signaling bandwidth (e.g., by adjusting the number of Physical Resource Blocks in scheduling) and pilots as the communication distance or channel coherence changes. Since both the optimal bandwidth $W^*$ and pilot ratio $\alpha^*$ changes with the cubic root of $L_c$, such adaptation is slow for typical wireless communication scenarios where user movement and channel dynamics are slow.}

Note that for reasonable range of beamforming gain $G$ and channel coherence $L_c$,  the optimized SNR $G\rho^*\simeq \sqrt[3]{4G/L_c}$ is  smaller than 1 and the corresponding spectral efficiency is around $0.1$~bits/s/Hz, which is substantially lower than the legacy cellular system. With practical range of available bandwidth ($\sim$GHz), when such low spectral efficiency transmission is optimal, the end user's rate will be in the range of hundreds Mbps or lower. For example, when a user's received signal power is low and the throughput maximizing bandwidth is in the range of tens to hundred MHz, the corresponding rate  will be in the range of a few Mbps.  Therefore the constant low spectral efficiency transmission is more suitable for  low-rate use cases.

\subsection{Throughput Improvement via Bandwidth  Optimization}

When user's received power is not high, using too much bandwidth will decrease the rate since the penalty from channel estimation becomes significant when SNR is too low. Therefore for users with low SNR, their rate will be increased if they use less bandwidth. The unused bandwidth can then be allocated to users with high SNR to improve their throughput. 

Let $0<g_1\leq g_2 \leq \cdots \leq g_k$ be the combined channel gain for a group of $k$ users served by the same base station. Under equal power and bandwidth allocation policy, their individual SNR will be
\[0< \frac{P_t g_1}{W_0N_0} \leq \frac{P_tg_2}{W_0N_0} \leq \cdots \leq \frac{P_tg_k}{W_0N_0},\]
 where $P_t$ is individual power budget and $W_0$ is the bandwidth. Under some fairness constraint, the joint bandwidth-power allocation problem can be formulated as a constrained optimization problem as follows
\begin{align}
 \max &\mbox{ a preselected objection function} \nonumber\\
\mbox{subject to }& P_i\geq 0, \ W_i\geq 0, \forall i, \nonumber\\
 & \sum_i P_i \leq kP_t, \ \sum_i W_i \leq kW_0, \label{eqn:PowerBW-opt} \\
& W_i = \arg\max_{W} R^*(W, \alpha|P_ig_i), \nonumber\\
& R(W_i,P_i)\geq R(W, P),\nonumber
\end{align} 
where the throughput maximizing bandwidth is obtained via the joint bandwidth and pilot optimization framework developed in 
 Sec.~\ref{sec:Opt-siso} and  Sec.~\ref{sec:Opt-mimo}, and the last constraint is a fairness constraint to ensure that no user's rate shall be decreased after participating the joint resource allocation. The objective function can be designed based on system level performance requirement.

In Fig.~\ref{fig:CDF_Rate_BW-Power-pooling} we evaluate the downlink throughput gain by two-user joint bandwidth-power allocation. The  user throughput  is evaluated by leveraging the CDFs of SINR generated for (a) a 28 GHz system~\cite{Ozge2015} with 228 meters inter site distance (ISD) equipped with $24\times1$ MIMO with transmit power $P_t=24$dBm and bandwidth $W_0=250$MHz budget for each user; (b) a 39 GHz system~\cite{Vook2016} with average ISD of 150 meters equipped with $32\times1$ MIMO subject to  68 dBm EIRP constraint and  $W_0=200$MHz per user bandwidth budget.  For convenience of simulation, the channel coherence is assumed to be $L_c=2.4\times10^4$ and  $L_c=3.2\times10^4$, respectively, to ensure an integer value of $L_c/G$. When we maximize the weak user's throughput, the 5-percentile rate can be improved by 10 times for the 28 GHz system and 9 times for the 39 GHz system, whereas the gain for medium and 95-percentile rate is negligible. However, if we choose to maximize the strong user's rate, the 5-percentile rate can be improved by roughly 5 times (by using less bandwidth) and the 95-percentile rate can be improved by 25\% and 30\% for the 28 GHz and 39 GHz system, respectively. The gain of the 95-percentile rate mainly comes from the exchange power for bandwidth with the weak users. Therefore a joint bandwidth-power allocation will make more efficient use of system resource.

\section{Conclusions}\label{sec:conclusion}

Given the abundant spectrum available in mmWave band, the challenge to provide high speed wireless access to users located a few hundreds meters away from the transmitter cannot be solved by simply adding more bandwidth. When channel estimation penalty is properly accounted for, as we have shown in this paper, the throughput gain by adding more spectrum is marginal\footnote{As shown in Fig.~\ref{fig:Opt_rate_W_1Watt_28GHz}, when the rate is close to its maximum, ten times bandwidth increase only buys less than two times rate gain.}, and it becomes counterproductive if we use too much bandwidth. 

Our analysis has revealed a surprising dependence of the optimal bandwidth and pilots ratio on the channel coherence length, which measures the average orthogonal symbols per each independent channel coefficient. When bandwidth is not limiting, for fixed channel coherence, the optimal bandwidth scales linearly with the received signal power, and the percentage of pilots is fixed whereas the absolute number of pilots grows linearly. Under reasonable assumptions, our analysis and numerical results have shown that the maximum beneficial bandwidth is below 100MHz for users located at a few hundreds meters away. We also show that, by joint allocation of power and bandwidth, under fairness constraint, the edge rate can be greatly increased. 

There are some limitations of our results. Our analysis is based on the IID block fading model, which is inherently a discrete model whereas the channel fading is continuous both in time and in frequency. The intra-block channel variation requires more pilot resource whereas the inter-block correlation has the potential to reduce the channel estimation penalty. {This I.I.D. block fading simplification ignores the inter-block correlation and discards the intra-block variation, and therefore represents the first-order approximation of continuous channel variation. A detailed analysis based on continuous channel variation is left to future work.} The impact of spatial correlation, {transmit/receiver array size, and the effective beamforming gain are yet to be investigated. A full treatment is out of the scope of this paper and a preliminary investigation onto this topic can be found in~\cite{Relay-Backhaul}.}

\appendices

\section{Derivation of Solution for the Continuous Valued Optimization}\label{app:opt_relax}

Since there is an $\E[\cdot]$, we have to first verify that  the condition of the \emph{Leibniz Integral Rule} holds (See \cite{NTNU-Note} for details) before we can move the partial derivative operator inside the expectation (which we will exercise later). 
That is, denoting 
\[f(W,\alpha,|h|)= \log(1+ \frac{\alpha L_c|h|^2P_r^2/N_0^2}{ W^2 + (1+\alpha L_c)WP_r/N_0})p_h(|h|),\]
where $p_h(|h|)$ is the probability density distribution (pdf) of the channel amplitude $|h|$, following~\cite[Theorem 5]{CUHK-Note}, we need to show that 
\begin{enumerate}
	\item $f(W,\alpha,|h|)$,  ${\partial f}/{\partial W}$, and ${\partial f}/{\partial \alpha}$ are continuous over $[B_c, W_{\mbox{max}}]\times[\frac{1}{L_c}, 1]\times[0,\infty)$;
	\item $\int_0^{\infty} f(W,\alpha, x) dx$, $\int_0^{\infty} {\partial f}/{\partial W} dx$, and $\int_0^{\infty} {\partial f}/{\partial \alpha} dx$ are uniformly convergent. 
\end{enumerate}

The first condition can be satisfied by assuming $p_h(|h|)$, the pdf of the small scale fading,  is continuous, which is well justified in most relevant communication scenarios.
To establish the second condition, we need to resort to the \emph{Lebesgue Dominated Convergence Theorem (LDCT)} by showing that for any given $\epsilon>0$, we can find $x_0>0$ such that for all $x_1, x_2>x_0$, we have
\begin{align}
\left|\int_{x_1}^{x_2} f(W,\alpha, x) dx\right|{<}\epsilon;  \left|\int_{x_1}^{x_2} \frac{\partial f}{\partial W} dx\right|{<}\epsilon;   \left|\int_{x_1}^{x_2} \frac{\partial f}{\partial \alpha} dx\right|{<}\epsilon. \label{eqn:D-E-exchange}
\end{align}

Fortunately, $|h|$ is the normalized channel amplitude representing the small scale fading, and its pdf $p_h(|h|)$ is generally regarded as continuous with infinite long tail of negligible probability. Since $\E[|h|^2]=1$, we can find a sufficiently large $x_0$ such that all the three inequalities in \eqref{eqn:D-E-exchange} hold. Therefore we conclude that we can exchange the order of partial derivation and expectation under the aforementioned assumptions on $|h|$.

We now take partial derivation of $R(W,\alpha)$ defined in \eqref{eqn:Rwa} with respect to $W$ and $\alpha$, respectively, shown in \eqref{eqn:dRdW} and \eqref{eqn:dRda} on the top of the next page,
where $\log_2(e)$ comes from the base substitution $\log_2(\cdot)\to\log(\cdot)$.  
\begin{figure*}
\begin{align}
\frac{\partial R}{\partial W} = & (1{-}\alpha)\log_2(e)\E\left[\log(1+ \frac{\alpha L_c|h|^2P_r^2/N_0^2}{ W^2 + (1+\alpha L_c)WP_r/N_0})\right] \label{eqn:dRdW}\\
&  +   (1{-}\alpha)W\log_2(e)\left( \E\left[\frac{2W+(1+\alpha L_c)P_r/N_0}{W^2 {+} (1{+}\alpha L_c)WP_r/N_0 {+} \alpha L_c |h|^2 P_r^2/N_0^2 } \right] {-} \frac{2W + (1{+}\alpha L_c)P_r/N_0}{W^2 {+} (1{+}\alpha L_c)WP_r/N_0}\right),  \nonumber\\
\frac{\partial R}{\partial \alpha} = & -W\log_2(e)\E\left[\log(1+ \frac{\alpha L_c|h|^2P_r^2/N_0^2}{ W^2 + (1+\alpha L_c)WP_r/N_0})\right] \label{eqn:dRda}\\
&  +  (1{-}\alpha)W\log_2(e)\left(\E\left[\frac{ WL_cP_r/N_0 +   L_c |h|^2P_r^2/N_0^2}{W^2 {+} (1{+}\alpha L_c)WP_r/N_0 {+} \alpha L_c |h|^2 P_r^2/N_0^2 }  \right] {-} \frac{ WL_c P_r/N_0 }{W^2 {+} (1{+}\alpha L_c)WP_r/N_0 } \right),  \nonumber
\end{align}
\hrulefill
\vspace*{2pt}
\end{figure*}
Setting ${\partial R}/{\partial W}=0$ and ${\partial R}/{\partial \alpha}=0$, and substituting $\rho=P_r/(N_0W)$  defined in \eqref{eqn:siso-SNR} to simplify the equations, we obtain
\begin{align}
& \E\left[\log(1+ \frac{\alpha L_c\rho^2|h|^2}{ 1 + (1+\alpha L_c)\rho})\right] \label{eqn:partial-A1} \\
 & = \frac{2+(1+\alpha L_c)\rho}{1 + (1+\alpha L_c)\rho} - \E\left[\frac{2+(1+\alpha L_c)\rho}{1 + (1+\alpha L_c)\rho + \alpha L_c \rho^2 |h|^2} \right], \nonumber\\
& \E\left[\log(1+ \frac{\alpha L_c\rho^2|h|^2}{ 1 + (1+\alpha L_c)\rho})\right] \label{eqn:partial-A2}\\
& =  
\E\left[\frac{ (1{-}\alpha)L_c\rho + (1{-}\alpha)L_c\rho^2 |h|^2 }{1 + (1+\alpha L_c)\rho + \alpha L_c \rho^2 |h|^2} \right] - \frac{ (1-\alpha)L_c\rho}{1 + (1+\alpha L_c)\rho}.  \nonumber
\end{align}

Combining \eqref{eqn:partial-A1} and \eqref{eqn:partial-A2}, we get  
\begin{align}
\frac{1+(1-\alpha)L_c\rho}{1 + (1+\alpha L_c)\rho} & = \E\left[\frac{1+(1-\alpha)L_c\rho + (1-2\alpha) L_c \rho^2 |h|^2}{1 + (1+\alpha L_c)\rho + \alpha L_c \rho^2 |h|^2}  \right]. \label{eqn:partial-A3} 
\end{align}
We first multiply both sides of \eqref{eqn:partial-A3} by $\alpha$ and then subtract $(1-2\alpha)$, which results in
\begin{align}
\frac{\alpha^2L_c\rho{-}(1{-}2\alpha)(1{+}\rho){+}\alpha}{1 + (1+\alpha L_c)\rho} = \E\left[\frac{\alpha^2L_c\rho{-}(1{-}2\alpha)(1{+}\rho){+}\alpha}{1 {+} (1{+}\alpha L_c)\rho {+} \alpha L_c \rho^2 |h|^2}  \right].\label{eqn:partial-A4} 
\end{align}
Since $|h|>0$ with $\E[|h|^2]=1$, we can see that \eqref{eqn:partial-A4} holds if and only if 
\begin{align}
\alpha^2L_c\rho-(1-2\alpha)(1+\rho)+\alpha = 0, \label{eqn:partial-A5} 
\end{align}
which can be written in two equivalent forms
\begin{align}
\alpha^2L_c\rho +\alpha(3+2\rho) -(1+\rho) = 0, \label{eqn:partial-A6} \\
\rho(\alpha^2L_c + 2\alpha -1) = 1-3\alpha. \label{eqn:partial-A7} 
\end{align}

\section{Algorithms to Determine $\rho^*$ and $\alpha^*$}\label{app:opt_num}
Since $\rho>0$, we can obtain from  \eqref{eqn:partial-A6} that
\begin{align}
\alpha^*  & = \frac{\sqrt{(\frac{3}{2}{+}\rho^*)^2 +(1{+}\rho^*)\rho^*L_c} - (\frac{3}{2}{+}\rho^*)}{\rho^*L_c} \nonumber\\
& =\frac{1+\rho^*}{\sqrt{(\frac{3}{2}{+}\rho^*)^2 +(1{+}\rho^*)\rho^*L_c} + \frac{3}{2}{+}\rho^*}, 
  \label{eqn:alpha_rho}
\end{align}
and from  \eqref{eqn:partial-A7}  that
\begin{align}
\left\{\begin{array}{cl}
\frac{1}{3} <\alpha^* <\frac{1}{1+\sqrt{1+L_c}}, & L_c<3,\\
\alpha^*=3, & L_c=3,\\
\frac{1}{1+\sqrt{1+L_c}} <\alpha^* < \frac{1}{3}, & L_c>3.
\end{array}\right.
\end{align}
 
Now we can either first substitute $\alpha(\rho)$ from \eqref{eqn:alpha_rho} into \eqref{eqn:partial-A1} to search for the throughput maximizing $\rho^*$ using the classical bisection method, and then use \eqref{eqn:alpha_rho} to determine $\alpha^*$, or first resort to the substitution \eqref{eqn:partial-A7} into \eqref{eqn:partial-A1} to search for the throughput maximizing $\alpha^*$ and then determine $\rho^*$ from \eqref{eqn:partial-A7}.  

Alternatively, we can substitute either $\alpha(\rho)$ defined in \eqref{eqn:alpha_rho} or  $\rho(\alpha)$ defined in \eqref{eqn:partial-A7} into the rate maximizing optimization problem \eqref{eqn:Opt_orig}, which can be efficiently solved by the bisection method.

\section{Closed-Form Approximation of $\rho^*$ and $\alpha^*$}\label{app:opt_approximation}

As channel coherence $L_c$ grows to infinity,  we have  $W^*\to\infty$ and $\alpha^*\to 0$, and the continuous-valued relaxation becomes plausible.  
We can approximate \eqref{eqn:partial-A1} by first applying $\E[\cdot]$ directly onto $|h|^2$ and then substituting $\log(1+x)\simeq x-x^2/2$, which leads to
\begin{align}
&\frac{\alpha L_c\rho^2}{ 1 + (1+\alpha L_c)\rho} - \frac{(\alpha L_c\rho^2)^2}{ 2(1 + (1+\alpha L_c)\rho)^2}\nonumber\\
&  = \frac{2+(1+\alpha L_c)\rho}{1 + (1+\alpha L_c)\rho} -  \frac{2+(1+\alpha L_c)\rho}{1 + (1+\alpha L_c)\rho + \alpha L_c \rho^2 }. \label{eqn:log_appr}  
\end{align}
Note that, with $\E[|h|^2]=1$,    
the relaxation to obtain \eqref{eqn:log_appr} shall introduce negligible impairment in the closed-form approximation as the throughput is maximized in the wideband regime where the SNR is low.

For low SNR, we can approximate \eqref{eqn:alpha_rho} by  
\begin{align}
\alpha & \simeq \frac{\sqrt{(1{+}\rho)\rho L_c} - \frac{3}{2}}{\rho L_c} \nonumber \\
 & \simeq (\rho L_c)^{-1/2}(1+\frac{1}{2}\rho -\frac{3}{2}(\rho L_c)^{-1/2}). \label{eqn:alpha_appr}
\end{align}
Substituting \eqref{eqn:alpha_appr} into \eqref{eqn:log_appr} and throwing away some smaller terms to simplify the derivation,  
we  obtain
\begin{align}
\rho^* & = 2(2L_c)^{-\frac{1}{3}} + O(L_c^{-\frac{2}{3}}) = \sqrt[3]{4/L_c} + O(L_c^{-\frac{2}{3}}), \label{eqn:rho_appr}\\
\alpha^* & =( {2L_c})^{-\frac{1}{3}} + O(L_c^{-\frac{2}{3}}) = \sqrt[3]{1/(2L_c)} + O(L_c^{-\frac{2}{3}}). \label{eqn:alpha_appr2}
\end{align}
To retain better approximation to  $\rho^*$ and $\alpha^*$ against the numerical solution proposed in Appendix~\ref{app:opt_num}, the following empirical approximation can be used ({at the same error level of $O(L_c^{-\frac{2}{3}})$})
\begin{align}
\rho^* & \simeq 2({2L_c})^{-\frac{1}{3}} + \frac{3}{2}(2L_c)^{-\frac{2}{3}}, \label{eqn:rho_heur}\\
\alpha^* & \simeq ({2L_c})^{-\frac{1}{3}} + \frac{5}{8}(2L_c)^{-\frac{2}{3}}. \label{eqn:alpha_heur}
\end{align}

\section*{Acknowledgment}

The authors would like to thank D. Chizhik, S. Venkatesan, and X. Shang for many helpful discussions.

\begin{IEEEbiography}[{\includegraphics[width=1.05in,height=1.1in, clip,keepaspectratio]{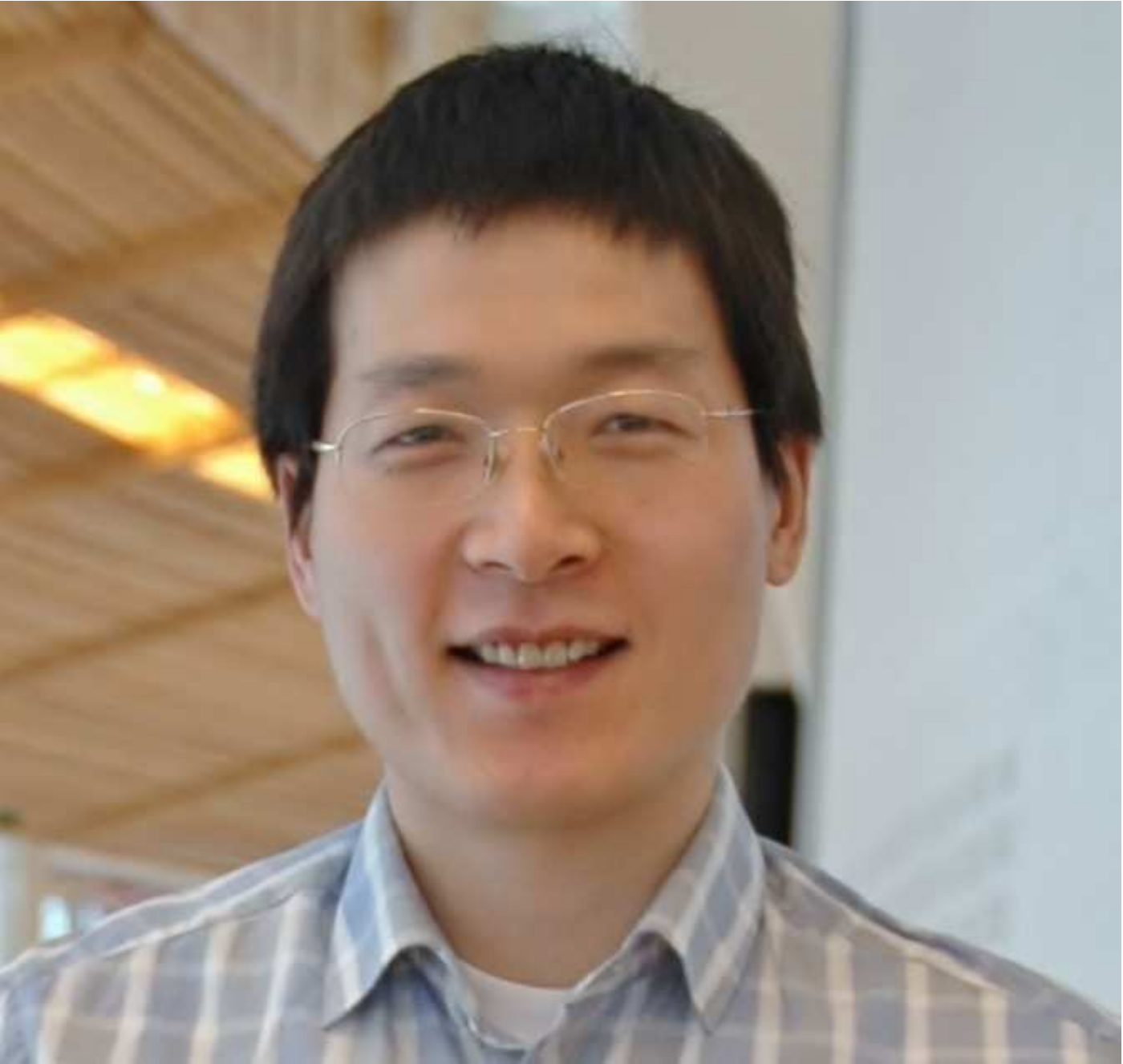}}]{Jinfeng Du} (S’07-M'13) received his B.Eng. degree in electronic information engineering from the University of
Science and Technology of China (USTC), Hefei, China, and the M.Sc., Tekn. Lic., and Ph.D. degrees from the Royal Institute of Technology (KTH), Stockholm, Sweden. He was a postdoctoral researcher at the Massachusetts Institute of Technology (MIT), Cambridge, MA, from 2013 to 2015, after which he joined Bell Labs at Crawford Hill, Holmdel, NJ, where he is currently a Member of Technical Staff. His research interests are in the general area of wireless communications, communication theory,  information theory,  and wireless networks. Dr.~Du received the Best Paper Award from IC-WCSP in October 2010, and his paper was elected as one of the ``Best 50 Papers'' in IEEE GLOBECOM 2014. He received the prestigious ``Hans Werth\'en Grant''  from the
Royal Swedish Academy of Engineering Science (IVA) in 2011, the ``Chinese Government Award for
Outstanding Self-Financed Students Abroad'' in 2012, and  the ``International PostDoc'' grant from
the Swedish Research Council in 2013. He also received three grants from the Ericsson Research Foundation.
\end{IEEEbiography}

\begin{IEEEbiography}[{\includegraphics[width=1in,height=1.1in, clip,keepaspectratio]{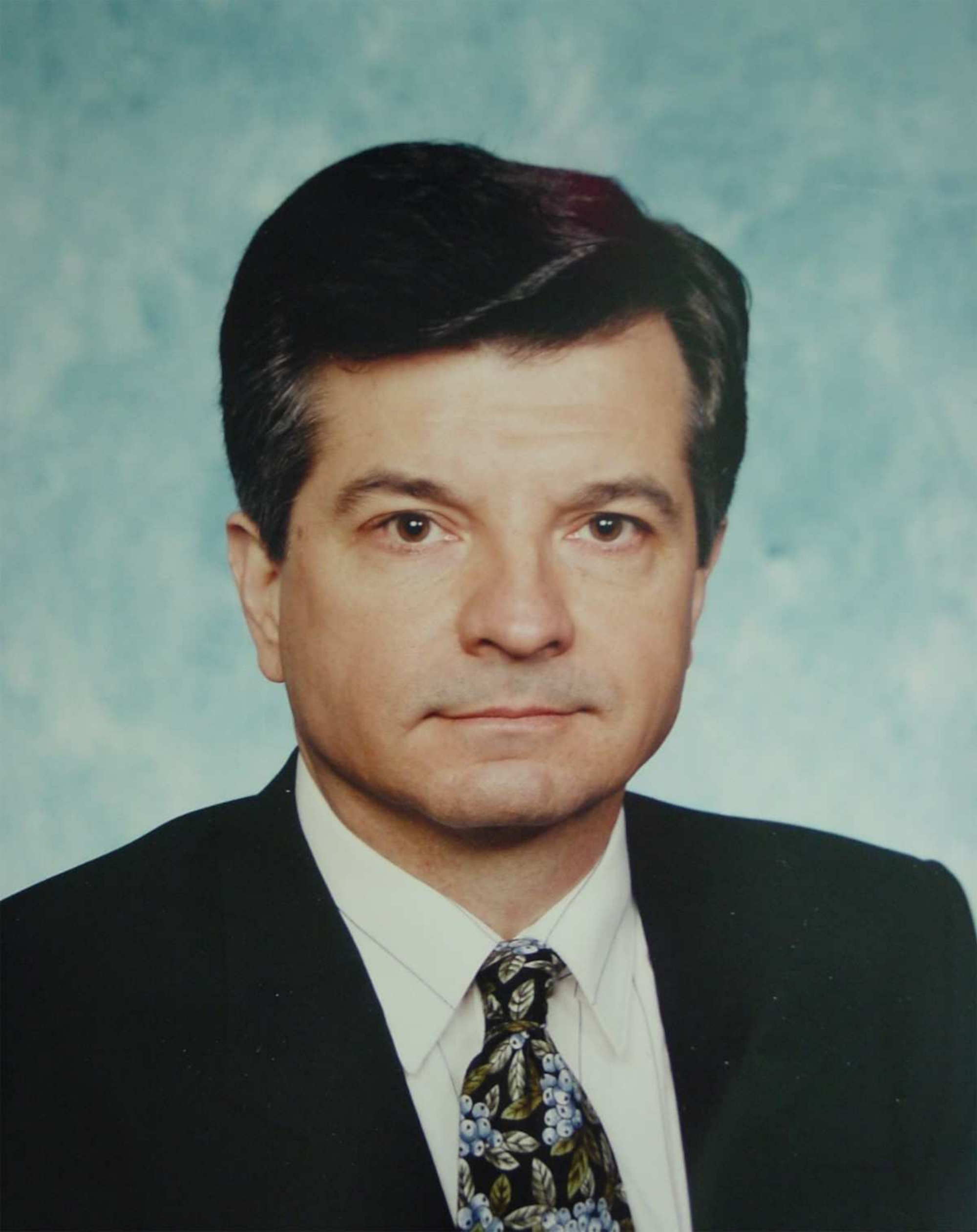}}]{Reinaldo A. Valenzuela} (M'85–SM'89–F'99) received the B.Sc. degree from the University of Chile, Santiago, Chile, and the Ph.D. degree from the Imperial College London, London, U.K. He is currently the Director of the Communication Theory Department and a Distinguished Member of Technical Staff with Bell Laboratories, Crawford Hill, NJ, USA. He is currently engaged in propagation measurements and models, MIMO/space time systems achieving high capacities using transmit and receive antenna arrays, HetNets, small cells and next generation air interface techniques and architectures. 

Dr. Valenzuela is a member of the US National Academy of Engineering and a Fellow of the IEEE. He is a Bell Labs Fellow, a WWRF Fellow, and a Fulbright Senior Specialist. He was a recipient of the 2010 IEEE Eric E. Sumner Award, the 2014 IEEE CTTC Technical Achievement Award, and the 2015 IEEE VTS Avant Garde Award. He has published 190 papers and 44 patents. He has over 26,000 Google Scholar citations and is a 'Highly Cited Author' In Thomson ISI.
\end{IEEEbiography} 

\end{document}